\documentclass[journal=gmj]{CUP-JNL-DTM}%

\usepackage{graphicx}
\usepackage{multicol,multirow}
\usepackage{amsmath,amssymb,amsfonts}
\usepackage{mathrsfs}
\usepackage{amsthm}
\usepackage{rotating}
\usepackage{appendix}
\usepackage{ifpdf}
\usepackage[T1]{fontenc}
\usepackage{newtxtext}
\usepackage{newtxmath}
\usepackage{textcomp}
\usepackage{xcolor}
\usepackage{lipsum}
\usepackage[colorlinks,allcolors=blue]{hyperref}
\usepackage{enumitem}
\usepackage{needspace}
\usepackage{longtable}
\usepackage{booktabs}
\usepackage{array}
\usepackage{ragged2e} 

\theoremstyle{definition}

\newcolumntype{L}[1]{>{\RaggedRight\arraybackslash}p{#1}}

\articletype{PREPRINT MANUSCRIPT}
\jyear{2026}
\begin{document}

\begin{Frontmatter}

\title[Reframing Population-Adjusted Indirect Comparisons as a Transportability Problem: An Estimand-Based Perspective and Implications for Health Technology Assessment]{Reframing Population-Adjusted Indirect Comparisons as a Transportability Problem: An Estimand-Based Perspective and Implications for Health Technology Assessment}

\author[1]{Conor Chandler}
\author[2]{Jack Ishak}

\authormark{Chandler \textit{et al}.}

\address[1]{\orgdiv{PPD}, \orgname{Thermo Fisher Scientific}, \orgaddress{\city{Waltham}, \state{MA},  \country{United States}}. \email{conor.chandler@thermofisher.com}}

\address[2]{\orgdiv{PPD}, \orgname{Thermo Fisher Scientific}, \orgaddress{\city{Montreal}, \state{Quebec},  \country{Canada}}}

\authormark{Chandler et al.}


\abstract{Population-adjusted indirect comparisons (PAICs) are widely used to synthesize evidence when randomized controlled trials enroll different patient populations and head-to-head comparisons are unavailable. Although PAICs adjust for observed population differences across trials, adjustment alone does not ensure that estimated effects are transportable to decision-relevant populations for health technology assessment (HTA).

\hspace*{4mm} Pairwise PAICs typically identify effects defined in the comparator
population, and applying these estimates to other populations entails an
additional, often implicit, transport step requiring further
assumptions. This has direct implications for HTA, where PAIC-derived
relative effects are routinely applied within cost-effectiveness models
defined for different target populations. In this paper, we formalize
this process from an estimand-based perspective and examine when such
transport is valid. We distinguish conditional and marginal estimands
and show that direct transportability depends jointly on the type of
effect modification, estimand, collapsibility, and alignment between the
scale of effect modification and the effect measure.

\hspace*{4mm} Using illustrative examples across continuous, binary, and time-to-event
outcomes, we demonstrate that even when effect modifiers are shared
across treatments, marginal effects are generally population-dependent
for commonly used non-collapsible measures, including hazard ratios and
odds ratios. Conversely, collapsible and conditional effects defined on
the linear predictor scale exhibit more favorable transportability
properties.

\hspace*{4mm} Our results clarify when applying PAIC-derived treatment effects to
desired target populations is justified, when doing so requires
additional assumptions, and when PAIC estimates should instead be
interpreted as population-specific rather than decision-relevant,
supporting more transparent and principled use of indirect evidence in
HTA and related decision-making contexts.}
\end{Frontmatter}

\hypertarget{introduction}{%
\section{Introduction}\label{introduction}}

Health technology assessment (HTA) agencies make decisions that
determine access to new interventions across diverse patient populations
and real-world healthcare settings.\textsuperscript{1-3} Randomized
controlled trials (RCTs) are pivotal in HTA, as they are regarded as the
gold standard for establishing causal effects.\textsuperscript{4} While
RCTs typically achieve strong internal validity, their external
validity---the degree to which findings can be meaningfully applied
beyond the study sample---may be limited.\textsuperscript{5-7} RCTs are
conducted under controlled conditions and often enroll narrowly defined
populations based on strict inclusion and exclusion
criteria.\textsuperscript{8} Consequently, treatment effects observed in
RCTs may not directly reflect outcomes in broader, more heterogeneous
patient populations encountered in routine clinical practice---the
populations most relevant to HTA
decision-making.\textsuperscript{5,9-11}

External validity encompasses two interrelated but distinct concepts:
generalizability and transportability.\textsuperscript{12,13}
Generalizability is defined by the extent to which a study sample is
representative of a target population.\textsuperscript{12,13}
Transportability refers to the ability to transfer a causal effect
learned in one population to a different external target population
(e.g., another trial's population or a real-world population) that may
differ in key characteristics such as baseline risk or distributions of
effect modifiers.\textsuperscript{9,12-17} In the HTA context,
transportability is the key consideration, since evidence must often be
extended from study populations to the real-world or jurisdictional
populations for which coverage and reimbursement decisions are
made.\textsuperscript{11,18}

Transportability issues are especially prominent in indirect treatment
comparisons (ITCs), which are widely used in HTA when no head-to-head
RCTs directly compare interventions of interest.\textsuperscript{19,20}
ITCs integrate and contrast evidence from different trials or data
sources to estimate relative treatment effects. Conventional ITC
methods, including Bucher adjustments and network meta-analyses (NMAs),
assume that all included studies are exchangeable, meaning that they
represent samples from a common underlying
population.\textsuperscript{21-24} Under this assumption, relative
treatment effects are invariant across studies and can be applied
without further adjustment. However, in practice, the populations
enrolled in different RCTs often differ meaningfully in baseline
characteristics and treatment effect modifiers, violating the
exchangeability assumption and limiting the validity of unadjusted
indirect comparisons.\textsuperscript{22,23,25,26}

To address these challenges, population-adjusted indirect comparisons
(PAICs)---such as the matching-adjusted indirect comparison (MAIC) and
simulated treatment comparison (STC)---have been
developed.\textsuperscript{19,27-32} These methods aim to improve
comparability across studies by adjusting for differences in observed
effect modifiers between study populations, typically leveraging
individual-patient level data (IPD) from at least one trial and
aggregate data from comparator studies. When correctly specified, they
can enhance the relevance of treatment effect estimates for HTA
decision-making.\textsuperscript{19,27,33} However, the use of PAICs
introduces important methodological and practical challenges. These
analyses rely on strong, often unverifiable assumptions---most notably,
that all relevant effect modifiers have been measured and that the
relationships between effect modifiers and treatment effects are
consistent across treatments.\textsuperscript{19,27-29,34} Moreover,
because IPD are rarely available for all relevant studies, it is
typically not possible to fully test the underlying assumptions or
assess model adequacy, which presents unique challenges to ensuring
transportable treatment effects.

Importantly, even when standard PAIC assumptions appear plausible,
uncertainty may remain about whether---and under what
conditions---PAIC-derived treatment effects can be validly applied to
the populations and settings relevant to HTA.\textsuperscript{35} In
practice, pairwise PAICs are often applied in a way that involves a
two-step transport process: first, treatment effects are conditionally
transported to the comparator trial population, and second, the
resulting indirect comparison is applied to a different
decision-relevant population. The validity of this second step is rarely
made explicit and largely depends on the transportability properties of
the estimand. To justify this second step, current HTA guidance
typically invokes the shared effect modifier assumption (SEMA). However,
SEMA alone is not sufficient to guarantee direct transportability (i.e., invariance of the treatment effect across populations, formally defined in Section 3.1). We
show that direct transportability is a joint property of (i) the
estimand being targeted, (ii) SEMA, (iii) alignment between the scale of
effect modification and the effect measure, and (iv) the collapsibility
of the effect measure. As a result, marginal treatment
effects---particularly for commonly used non-collapsible measures such
as odds ratios and hazard ratios---are generally population-dependent
and not directly transportable, whereas conditional effects and certain
collapsible marginal effects exhibit more favorable transportability
properties.

In this paper, we reframe PAICs as transportability problems and examine
their implications from an estimand-based perspective. We first define
conditional and marginal estimands and explain how their population
dependence gives rise to the transportability problem. We then show how
transportability assumptions are embedded---explicitly or
implicitly---in commonly used ITC and PAIC methods, with particular
focus on the two-step process often arising in applications of pairwise
MAICs and STCs.\textsuperscript{9,12,13,18,36-41} Building on this
framework, we characterize the conditions under which PAIC-derived
relative effects are directly transportable rather than
population-specific. We also discuss practical considerations for
ensuring that comparative effectiveness evidence is applicable for HTA
decision making. Although we focus primarily on PAICs to motivate
concepts, the issues we describe arise whenever treatment effects are
transported across populations, including in standard NMAs and
cost-effectiveness models.

\hypertarget{causal-estimands-and-population-dependence-in-transportability-analyses}{%
\section{Causal Estimands and Population Dependence in Transportability
Analyses}\label{causal-estimands-and-population-dependence-in-transportability-analyses}}

To understand transportability in PAICs, it is essential to clarify what
treatment effect is being estimated (the estimand) and how it depends on
the population in which it is evaluated. A key distinction is between
population-average conditional and marginal effects, which differ in how
they aggregate outcomes over the covariate distribution and therefore
have different transportability properties. Because marginal effects
involve averaging after a (typically nonlinear) transformation, they are
typically population-dependent, whereas conditional effects may be
invariant in indirect comparisons under appropriate modeling
assumptions. Much of our discussion focuses on marginal effects, which
are most commonly used in HTA and are central to the transportability
challenges we examine in MAIC and STC.

Throughout this paper, we emphasize that transportability is a joint
property of the estimand and the modeling assumptions used to identify
it (e.g., SEMA). This section introduces the causal framework and
defines the estimands used throughout the paper. These concepts provide
the foundation for the transportability results that follow, where we
examine when treatment effects are invariant across populations and when
they are inherently population-specific.

\hypertarget{causal-framework}{%
\subsection{Causal Framework}\label{causal-framework}}

We adopt the Rubin Causal Model (RCM) to define causal treatment
effects. Let \(Y^{t}\) denote the potential outcome under treatment
\(t \in \{ A,B\}\).\textsuperscript{41-44} The causal estimand compares
the expected potential outcomes under treatments \(A\) and \(B\):

\begin{equation}
\psi\!\left(
\mathbb{E}\{Y^{A}\},
\mathbb{E}\{Y^{B}\}
\right)
\label{eq:1}
\end{equation}

where $\psi(\cdot)$ is a suitable contrast function measuring the
causal effect of \(B\) vs. \(A\). For instance, this may be expressed as
a difference $\mathbb{E}\{Y^{B}\}-\mathbb{E}\{Y^{A}\}$ or a ratio
$\mathbb{E}\{Y^{B}\}/\mathbb{E}\{Y^{A}\}$.

\hypertarget{population-average-estimands}{%
\subsection{Population-average
Estimands}\label{population-average-estimands}}

In comparative effectiveness research, the goal is often to estimate
population-average causal effects;\textsuperscript{19,35,38} that is,
the expected difference in outcomes between treatment \(A\) and
treatment \(B\) when applied to a specified population. These
population-level effects can be expressed in two closely related forms:
conditional effects and marginal effects.\textsuperscript{19,35,45}

\hypertarget{population-average-conditional-effects}{%
\subsubsection{Population-average Conditional
Effects}\label{population-average-conditional-effects}}

Population-average conditional effects average individual-level causal
contrasts over the covariate distribution \(X\), possibly after
transformation:

\begin{equation}
\Delta^{\mathrm{Cond}}
=
\mathbb{E}_{X}
\!\left[
h\!\left(\mathbb{E}\{Y^{B}\mid X=x\}\right)
-
h\!\left(\mathbb{E}\{Y^{A}\mid X=x\}\right)
\right]
\label{eq:cond_estimand}
\end{equation}

The function \(h( \cdot )\) represents a transformation defining the
effect measure of interest, which may be applied on the outcome scale or
on a transformed outcome scale (e.g., a log link when working with
ratios to transform them into an additive scale, or an identity link for
risk differences).

When the conditional expectation of the outcome is linear in the
covariates (i.e., effects are additive on the linear predictor scale),
the population-average conditional effect simplifies to:

\begin{equation}
\begin{aligned}
\Delta^{\mathrm{Cond}}
&=
\mathbb{E}_{X}\!\left\{
\mathbb{E}\!\left\{Y^{B}\mid X=x\right\}
-
\mathbb{E}\!\left\{Y^{A}\mid X=x\right\}
\right\}
\\
&=
\mathbb{E}_{X}\!\left\{
\mathbb{E}\!\left\{Y^{B}\mid X=x\right\}
\right\}
-
\mathbb{E}_{X}\!\left\{
\mathbb{E}\!\left\{Y^{A}\mid X=x\right\}
\right\}
\\
&=
\mathbb{E}\!\left\{Y^{B}\mid X=\bar{x}\right\}
-
\mathbb{E}\!\left\{Y^{A}\mid X=\bar{x}\right\}.
\end{aligned}
\label{eq:cond_effect}
\end{equation}

where averaging over \(X\) is equivalent to evaluating the model at the
mean covariate values under linearity assumptions. In such situations,
treatment effects are less sensitive to differences in covariate
distributions across populations, which can facilitate transportability
under these linearity assumptions.

\hypertarget{marginal-effects}{%
\subsubsection{Marginal Effects}\label{marginal-effects}}

Marginal effects quantify the change in average potential outcomes
across the whole population if everyone were to receive treatment \(B\)
versus treatment \(A\):

\begin{equation}
\Delta^{\mathrm{Marg}}
=
h\!\left(\mathbb{E}_{X}\{Y^{B}\}\right)
-
h\!\left(\mathbb{E}_{X}\{Y^{A}\}\right).
\label{eq:marg_effect}
\end{equation}

The key distinction is where the expectation over \(X\) occurs relative
to the transformation \(h( \cdot )\). For conditional effects, the
transformation is applied before averaging; for marginal effects,
averaging occurs first. Therefore, marginal effects can depend on the
full covariate distribution in ways that differ from conditional
effects.

\hypertarget{dependence-on-the-population-of-study}{%
\subsubsection{Dependence on the Population of
Study}\label{dependence-on-the-population-of-study}}

Both conditional and marginal estimands are population-dependent, but
they depend on the population in different ways. This overview focuses
on marginal treatment effects, as the methods and HTA practices that
motivate our central argument most often target marginal estimands.
These concepts generalize straightforwardly to conditional effects.

To make population dependence explicit, let \(P_{1}\) denote a source
(index) trial population and \(P_{2}\) denote an external target
population. The marginal effect in population \(P \in \{ P_{1},P_{2}\}\)
is

\begin{equation}
\begin{aligned}
\Delta_{P}^{\mathrm{Marg}}
&=
\psi\!\left(
\mathbb{E}_{P}\{Y^{B}\},
\mathbb{E}_{P}\{Y^{A}\}
\right)
\\
&=
h\!\left(
\mathbb{E}_{P}\{Y^{B}\}
\right)
-
h\!\left(
\mathbb{E}_{P}\{Y^{A}\}
\right).
\end{aligned}
\label{eq:marg_p}
\end{equation}

Here, \(\mathbb{E}_{P}\!\left\{ Y^{t} \right\}\) denotes the expected
potential outcome under treatment \(t\) in population \(P\).

In general, $\Delta_{P_{1}}^{\mathrm{Marg}} \neq \Delta_{P_{2}}^{\mathrm{Marg}}$ when
covariate distributions differ across populations, because marginal
effects are defined by averaging potential outcomes over the population,
which depends on both the covariate distribution and how treatment
effects vary with covariates (i.e., effect modification). As a result, a
treatment effect estimated in one population may not apply in another.
Importantly, even after adjustment with PAICs, these methods may still
target estimands that vary across populations. Understanding when such
estimands are invariant---which facilitates direct transportability to
decision-relevant populations in HTA---is a key focus of the remainder
of the paper.

\hypertarget{when-and-how-are-treatment-effects-transportable-across-populations}{%
\section{When and How Are Treatment Effects Transportable Across
Populations?}\label{when-and-how-are-treatment-effects-transportable-across-populations}}

In HTA, evidence generated in clinical trial populations is typically
applied to different populations for which reimbursement and coverage
decisions must be made.\textsuperscript{18,19,45} However, when
populations differ in their covariate distributions, treatment effects
may also differ, making transportability a non-trivial assumption.
Transportability is concerned with whether a causal effect learned in
one population can be validly applied to a different target
population.\textsuperscript{9,13,15} We introduce the notation
\(\Delta_{P_{1} \rightarrow P_{2}}\) to explicitly distinguish the
population in which evidence is learned (\(P_{1}\)) from the population
in which the effect is evaluated (\(P_{2}\)); when \(P_{1} = P_{2}\),
this reduces to the usual population-specific estimand \(\Delta_{P}\).

\hypertarget{direct-transportability-unadjusted}{%
\subsection{Direct Transportability
(Unadjusted)}\label{direct-transportability-unadjusted}}

A treatment effect estimand is directly transportable from a source
population \(P_{1}\) to a target population \(P_{2}\) if the causal
effect evaluated in \(P_{1}\) is equal to the causal effect evaluated in
\(P_{2}\).\textsuperscript{39} Direct transportability means that the
causal effect is invariant to changes in the population.

\begin{equation}
\Delta_{P_{1} \to P_{2}}^{\mathrm{Marg}}
=
\Delta_{P_{1}}^{\mathrm{Marg}}
=
\Delta_{P_{2}}^{\mathrm{Marg}}
\end{equation}

The equality in Equation (6) may hold, for example, when there is no
effect modification by covariates whose distributions differ between
populations, or when the relevant effect-modifying covariates have the
same distribution in both populations. In many applied settings,
however, these conditions are unlikely to be met. Differences in
baseline characteristics---such as age, disease severity, or biomarker
profiles---can induce treatment effect heterogeneity, making unadjusted
transport of effects across populations invalid.

\hypertarget{conditional-transportability-population-adjusted}{%
\subsection{Conditional Transportability
(Population-adjusted)}\label{conditional-transportability-population-adjusted}}

When direct transportability does not hold, effects may still be
transportable conditionally. Conditional transportability is invoked
when differences in treatment effects across populations can be
explained by a set of observed covariates, denoted
\(X\).\textsuperscript{17} In this case, the transported marginal
treatment effect from population \(P_{1}\) to \(P_{2}\) is generally
defined as:

\begin{equation}
\Delta_{P_{1} \to P_{2}}^{\mathrm{Marg}}
=
\psi\left(
\mu_{P_{1} \to P_{2}}^{B},\,
\mu_{P_{1} \to P_{2}}^{A}
\right)
\end{equation}

where
\(\mu_{P_{1} \rightarrow P_{2}}^{t} : = \mathbb{E}_{P_{2}}\left\{ Y^{t} \right\}\)
represents the transported expected potential outcome under treatment
\(t \in \{ A,B\}\) in the target population \(P_{2}\), estimated using
information learned from the source population \(P_{1}\). Importantly,
conditional transportability does not imply that treatment effects are
invariant across populations; rather, it defines how to re-express
effects for a specific target population.

\hypertarget{modes-of-conditional-transportability-in-paics}{%
\subsubsection{Modes of Conditional Transportability in
PAICs}\label{modes-of-conditional-transportability-in-paics}}

There are generally two types of approaches used to estimate the
transported potential outcomes \(\mu_{P_{1} \rightarrow P_{2}}^{t}\): 1)
outcome regression modeling and 2) weighting
methods.\textsuperscript{13} These correspond to commonly used pairwise
PAIC methods such as STC (outcome regression) and MAIC
(weighting).\textsuperscript{27,28} Both methods adjust for differences
in \(X\) between populations and aim to estimate treatment effects in
the comparator study population \(P_{2}\) based on information from the
population \(P_{1}\) in the sponsor's study.

In outcome regression, the transported potential outcomes are derived by
modeling the conditional outcome in \(P_{1}\) and averaging predicted
outcomes over the covariate distribution of \(P_{2}\):

\begin{equation}
\mu_{P_{1} \to P_{2}}^{t}
=
\mathbb{E}_{X \sim P_{2}}\!\left[
\mathbb{E}_{P_{1}}\!\left\{ Y \middle| T = t, X \right\}
\right]
\end{equation}

where \(X\sim P_{2}\) indicates that \(X\) are distributed according to
the target population \(P_{2}\).\textsuperscript{46}

The outer expectation in Equation (8) is evaluated by integrating over
the covariate distribution in \(P_{2}\):

\begin{equation}
\mu_{P_{1} \rightarrow P_{2}}^{t}
= \int
\underbrace{\mathbb{E}_{P_{1}}\!\left( Y \mid T=t, X=x \right)}_{\text{Model for source}}
\,
\underbrace{f_{P_{2}}(x)}_{\text{\(X\) in target}}
\, dx
\end{equation}

where \(f_{P_{2}}(X)\) denotes the joint probability density function of
\(X\) in \(P_{2}\).\textsuperscript{9,47}

Under this representation, the outcome regression model for STC can be
written as

\begin{equation}
g\!\left(
\mathbb{E}\{Y^{t}\mid X\}
\right)
=
\eta_{t}(X).
\end{equation}

where $g(\cdot)$ denotes the link function and \(\eta_{t}(X)\) denotes the linear predictor under treatment
\(t\). In practice, the model is fitted to IPD from the \(P_{1}\) to
estimate the relationship between the outcome, treatment, and
covariates. The fitted model is then used to predict potential outcomes
under treatment \(t\) for individuals in \(P_{2}\), which are averaged
to obtain the transported outcomes and effects. The integral in Equation
(9) is typically approximated using Monte Carlo integration:

\begin{equation}
\mu_{P_{1}\rightarrow P_{2}}^{t}
\approx
\frac{1}{n_{2}}
\sum_{i=1}^{n_{2}}
g^{-1}\!\left(
\widehat{\eta}\!\left(T=t,\,X=x_{i2}\right)
\right).
\label{eq:transport_mean}
\end{equation}

Here, \(x_{i2}\) denotes the covariate vector for individual \(i\) in
\(P_{2}\), and \(g^{- 1}( \cdot )\) is the inverse link function. This
approach is commonly referred to as
g-computation.\textsuperscript{9,47-49} In a few special cases (Section
6.2), this integral simplifies to evaluating the model at the mean of
the covariates in \(P_{2}\), reflecting settings where effects may be
more easily transportable.

Multilevel network meta-regression (ML-NMR) is a hierarchical outcome
regression framework that integrates individual-level modeling, as in
STC, with the evidence-synthesis structure of NMA, allowing information
to be combined across studies and connected treatment contrasts in a
network. An important distinction between STC and ML-NMR concerns how
baseline risk is represented: STC relies on a single intercept anchored
to the source population, whereas ML-NMR allows study-specific
intercepts. This highlights that transportability depends not only on
adjustment for effect modifiers, but also on how baseline risk is
modeled---an aspect that differs across these two methods. We return to
this distinction in Section 9.2.

Alternatively, the transported potential outcomes can be estimated using
weighting methods, which aim to adjust the distribution of covariates in
\(P_{1}\) to resemble that of the target population
\(P_{2}\).\textsuperscript{15,50-52}

Under the weighting framework, the transported potential outcome under
treatment \(t\) is estimated as:

\begin{equation}
\mu_{P_{1} \to P_{2}}^{t}
=
\frac{
\mathbb{E}_{P_{1}}\!\left[
w_{P_{2}}(X)\,\mathbf{1}(T = t)\,Y
\right]
}{
\mathbb{E}_{P_{1}}\!\left[
w_{P_{2}}(X)\,\mathbf{1}(T = t)
\right]
}
\end{equation}

where \(w_{P_{2}}(X)\) are weights that align the source population with
the target covariate distribution. Weighting methods are particularly
useful when outcome models are complex or prone to misspecification but
can be sensitive to extreme weights when overlap between populations is
limited.\textsuperscript{50,51}

Doubly robust methods could also be employed to transport treatment
effects. These approaches combine both outcome regression modeling and
weighting-based adjustments to provide more reliable estimates of the
transported potential outcomes and effects. In particular, doubly robust
methods offer additional protection against model misspecification: the
estimator remains consistent if either the outcome model or the
weighting model is correctly specified, though not necessarily
both.\textsuperscript{53-56}

Equations (8) -- (12) highlight that MAIC and STC target estimands
defined with respect to the covariate distribution in the comparator
study. Section 4 considers the implications of subsequently applying these estimates to a different decision-relevant population.

\hypertarget{transportability-in-pairwise-paics-a-dual-challenge}{%
\section{Transportability in Pairwise PAICs: A Dual
Challenge}\label{transportability-in-pairwise-paics-a-dual-challenge}}

The transportability concepts introduced in Section 3 are particularly
consequential in the context of pairwise PAICs, where asymmetries in
data availability require treatment effects to be transported across
populations in a structured---but often implicit---manner. In practice,
this involves a sequence of transport steps rather than a single
adjustment. Although we illustrate transportability using pairwise
PAICs, many of the same concepts apply to other ITCs (e.g., NMAs) and
propagate forward into cost-effectiveness models that use these effect
estimates as inputs.

In anchored ITCs, IPD are typically available only for the index trial
comparing a new intervention \(B\) with a common comparator \(A\), while only
aggregate data are available for a comparator trial evaluating an
alternative intervention \(C\) versus \(A\) (Figure 1). As a result, pairwise
PAICs such as MAICs and STCs necessarily treat the index trial as the
source population and the comparator trial as the target population,
yielding relative treatment effect estimates defined in the comparator
population.\textsuperscript{19} This population, however, is often not the one relevant for decision making in HTA.

\vspace{5mm}
\noindent\textbf{Figure 1.} Example Network for Anchored PAIC Involving Two Studies

\includegraphics[scale=0.85]{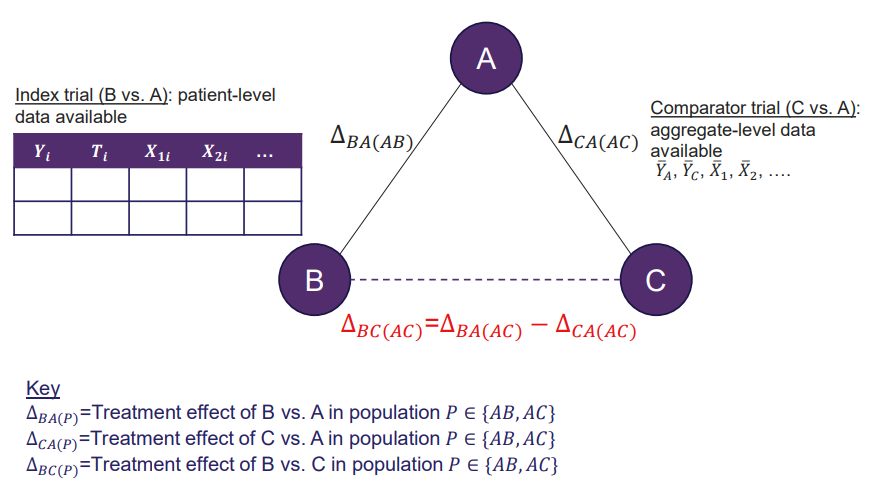}
\vspace{5mm}

In HTA applications, pairwise MAICs and STCs are typically applied in a
manner that entails two transport steps, each relying on distinct
transportability assumptions.

\vspace{1.75mm}
\noindent \textbf{Step 1: Conditional transport to the comparator population:}
\vspace{1.75mm}

In the first step, the treatment effect for \(B\) vs. \(A\),
\(\Delta_{BA}\), is conditionally transported from the index to the
comparator study population to estimate \(B\) vs. \(C\)
(\(\Delta_{BC}\)) in the comparator population (Section 3.2.1). MAICs achieve
this by reweighting index trial participants such that the weighted
distribution of effect modifiers matches the aggregate characteristics
reported for the comparator trial. For instance, if the comparator
population is composed of 67\% low-risk and 33\% high-risk patients, the
MAIC weighting ensures the sample from the index trial reflects the same
proportions. This adjustment results in \(\Delta_{BA(AC)}\), adjusted
effect for \(B\) vs. \(A\) in the comparator population. Then, the
adjusted effect for \(B\) vs. \(C\) in the comparator (AC) population is
estimated on the additive scale by comparing the adjusted effect for
\(B\) vs. \(A\) with the reported effect for \(C\) vs. \(A\):
\(\Delta_{BC(AC)} = \Delta_{BA(AC)} - \Delta_{CA}\). STCs instead use
outcome regression models fitted in the index trial to predict outcomes
under treatment \(B\) for covariate profiles representative of the
comparator population. This first step relies on the assumption of
conditional transportability; that is, that all relevant effect
modifiers for $B$ vs. $A$ have been correctly identified and adjusted for.

\vspace{1.75mm}
\noindent \textbf{Step 2: Implicit direct transport to the decision-relevant population:}
\vspace{1.75mm}

In HTA, the comparator trial population is rarely the population of
interest for decision-making.\textsuperscript{57} Instead, relative
treatment effects are typically applied within cost-effectiveness models
defined for the index trial population or for broader real-world
populations relevant to reimbursement decisions. Consequently, the
adjusted effect estimate from Step 1 (\(\Delta_{BC})\) is often applied
directly to a different target population---typically, the index
trial---implicitly assuming direct transportability of the
active-to-active effect, without being formally identified as a second
transport step in practice. For example, \(\Delta_{BC(AC)}\) may be an
adjusted hazard ratio for \(B\) vs. \(C\) estimated in the comparator
trial population. In cost-effectiveness models for HTA, this hazard
ratio is often applied to a baseline survival curve for \(B\) estimated
from the index trial, under the assumption that this baseline represents
the decision-relevant population, to generate a counterfactual survival
curve for \(C\) in the index population.\textsuperscript{58} Applying
\(\Delta_{BC(AC)}\) to a baseline curve from the index trial implicitly
assumes that the hazard ratio is directly transportable and valid in the
index population represented by that curve, i.e.,
\(\Delta_{BC(AC)} = \Delta_{BC(AB)}\). This type of workflow, in which
relative effects from evidence synthesis are combined with baseline
survival models in cost-effectiveness analysis, is well established in
HTA decision modeling.\textsuperscript{58,59} Analogous applications may
occur for other types of outcomes (e.g., binary) when PAIC-derived
effects are applied to baseline risks or outcomes from a
decision-relevant population;\textsuperscript{60} the consequences of
doing so differ by effect measure and are summarized in Table 1. National Institute for Health and Care Excellence (NICE) Decision Support Unit (DSU) Technical Support Document (TSD) 18 formalizes the role of SEMA in this second step, asserting that it ensures the ability to transfer inferences of
active-to-active indirect comparisons (i.e., \(\Delta_{BC}\)) based on
MAIC or STC between studies.\textsuperscript{57}

The two-step nature of how pairwise MAICs and STCs are applied in
practice creates a dual challenge (Figure 2). First-order transport
(Step 1) must achieve valid conditional transportability from the index
to the comparator population by correctly adjusting for all relevant
effect modifiers. Second-order transport (Step 2) then assumes direct
transportability of the resulting effect from the comparator to a
different target population, relying heavily on SEMA. Each stage
introduces its own potential sources of bias. If effect modifiers are
misspecified or incompletely reported, Step 1 yields a biased estimate
of \(\Delta_{BC(AC)}\). Step 2 relies on SEMA, but we demonstrate in
subsequent sections that this assumption may not be sufficient for
direct transportability and may introduce substantial bias. Much of the
confusion surrounding the interpretation of PAIC results in HTA arises
from failing to distinguish these two transport steps and from assuming
that successful Step 1 adjustment, together with SEMA, is sufficient to
justify Step 2. The second step is the central methodological challenge
examined in this paper.

\vspace{2mm}
\noindent\textbf{Figure 2.} Two-step Process to Transporting Effects in Pairwise MAICs and
STCs

\includegraphics[scale=0.85]{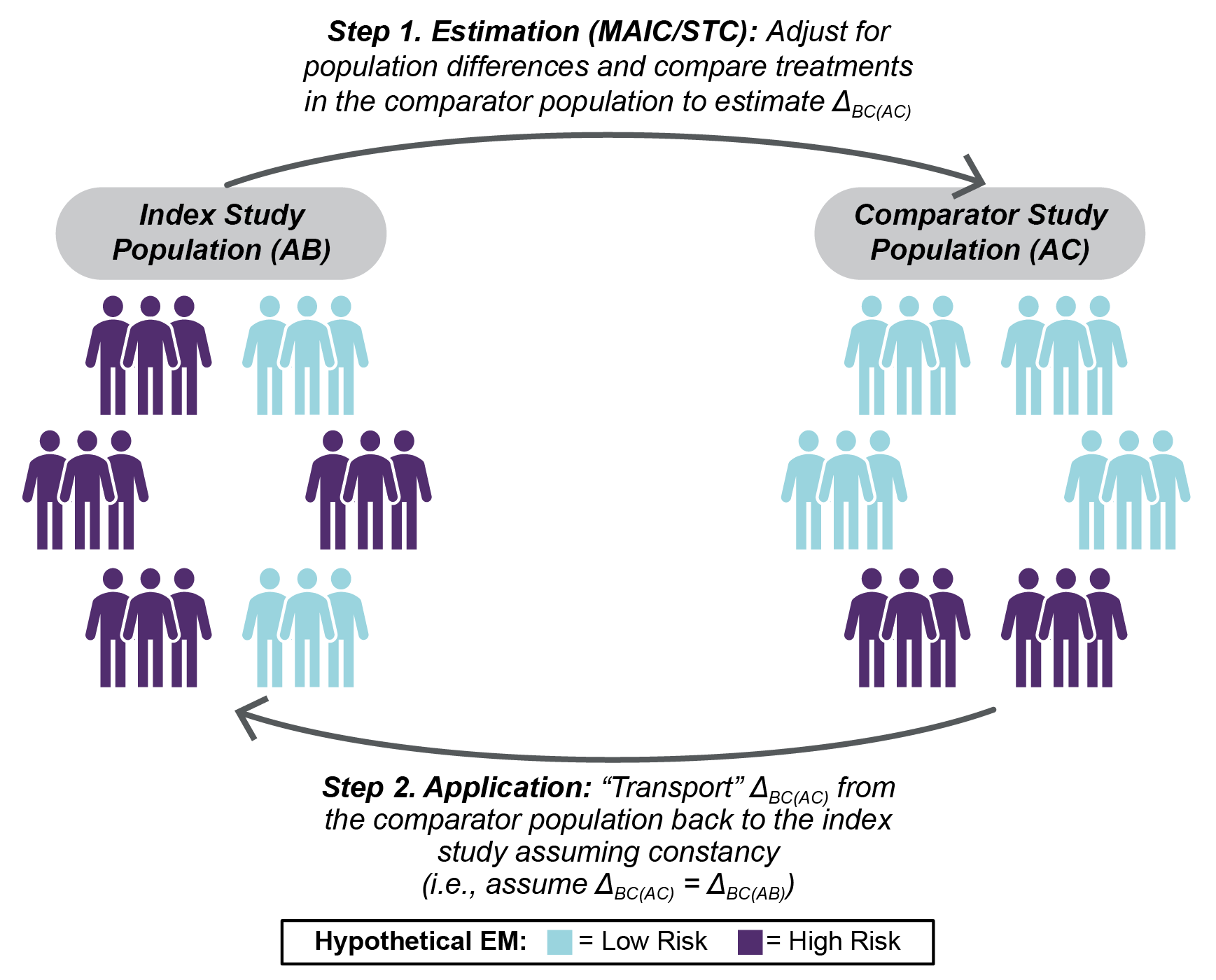}

{\footnotesize
\setlength{\baselineskip}{0.9\baselineskip}
\noindent Abbreviation: EM = effect modifier \par
\noindent Note: The index study population is shown as the target population for Step 2 for illustration; in practice, the target may be any decision-relevant population. \par
}

\hypertarget{central-role-of-the-shared-effect-modifier-assumption}{%
\section{Central Role of the Shared Effect Modifier
Assumption}\label{central-role-of-the-shared-effect-modifier-assumption}}

In PAICs, SEMA comprises two key conditions when comparing a set of
active treatments (e.g., \(B\) and \(C\)) relative to a common
comparator (\(A\)): (1) the same set of covariates function as effect
modifiers; and (2) the strength and direction of each effect modifier's
impact on the treatment effect are identical across all active
treatments.\textsuperscript{19,28,45} SEMA plays a pivotal role in
PAICs, as it is often invoked to facilitate transporting effects across
populations.\textsuperscript{45,57,61} When SEMA holds, current guidance
suggests that effect modifiers ``cancel out'' for active-to-active
indirect comparisons (e.g., \(B\) vs. \(C\)), thereby enabling the
direct transportability of \(\Delta_{BC}\) to any population, including
the index study.\textsuperscript{28,57,61} Historically, both MAICs and
STCs have relied on SEMA to justify transporting treatment effects for
active-to-active comparisons (e.g., \(B\) vs. \(C\)) from the comparator
population to the desired target population (e.g., index
study).\textsuperscript{28,57,61}

The importance of SEMA in the context of transportability in PAICs is
underscored in current guidance from key HTA bodies. NICE DSU TSD 18
states the following: \emph{``{[}\ldots{]} if the shared effect modifier
assumption holds for treatments B and C, then the estimated marginal
relative treatment effect (whether obtained using anchored or unanchored
MAIC/STC) will be applicable to any population.''\textsuperscript{57}}

Similarly, the HTA coordination group (CG), which is responsible for
overseeing the Joint Clinical Assessments (JCAs) aimed at evaluating the
clinical effectiveness of new health technologies across the European
Union (EU), supports NICE's position. The HTA CG has concluded that it
is feasible and appropriate to directly transport relative effect
estimates obtained from MAIC or STC to the desired target population
when SEMA is met, stating: \emph{``When this {[}SEMA{]} holds, the
relative effect between any pair of treatments in this set {[}of active
treatments{]} will be the same in any population, which means that
treatment effects obtained from population-adjusted indirect comparisons
{[}MAICs and STCs{]} can be transposed to the population of the source
{[}index{]} trial or indeed any other relevant
population''}\textsuperscript{61}

Despite its central role in HTA guidance, SEMA alone does not guarantee
direct transportability of marginal treatment
effects.\textsuperscript{45} In Sections 6 and 7 we provide both
theoretical arguments and present illustrative examples to clarify that,
in general, SEMA is not always sufficient to guarantee direct
transportability of \(\Delta_{BC}\). If SEMA is met and both effect
modification and effect measures are modeled on the linear predictor
scale (Section 6.3), then population-average conditional effects are
directly transportable. For marginal effects, only certain measures
(e.g., mean differences, risk ratio), which are said to be collapsible
(Section 6.2), are directly transportable if these assumptions are met;
however, marginal effects for measures that are non-collapsible,
including odds ratios and hazard ratios---which are the most common
measures compared in practice---are not directly transportable, even
when SEMA holds. Therefore, the current guidance\textsuperscript{57,61}
is not fully sufficient, as SEMA alone is not proper justification for
transporting effects in MAICs and STCs, as these methods typically
target marginal estimands rather than conditional estimands.

\hypertarget{beyond-sema-estimands-and-conceptual-foundations}{%
\section{Beyond SEMA: Estimands and Conceptual
Foundations}\label{beyond-sema-estimands-and-conceptual-foundations}}

Transportability in PAICs depends not only on SEMA, but also on the estimand of interest, collapsibility, and the correspondence between the scale of effect modification and the effect measure. These concepts collectively define the conditions under which relative effects are directly transportable across populations.

\hypertarget{type-of-target-estimand-conditional-vs.-marginal}{%
\subsection{Type of Target Estimand: Conditional vs.
Marginal}\label{type-of-target-estimand-conditional-vs.-marginal}}

As discussed in Section 2.2, the distinctions between conditional and
marginal estimands are central to PAICs, which seek to estimate
treatment effects that are valid in a target population when only
aggregate data are available from comparator trials. Different PAIC
approaches target different estimands.\textsuperscript{19,45} MAIC
targets marginal effects by construction. STC and ML-NMR can target
either population-average conditional or marginal effects, depending on
the model and post-processing steps. However, STCs are often constrained
in practice by the form of the effect measure reported in the comparator
study, which is typically marginal.\textsuperscript{19,45}

Sections 5, 6.2, and 6.3 examine how SEMA, collapsibility, and the
alignment between the model scale and the desired effect measure
influence the transportability properties of the estimand being
targeted. When both the model link \(g( \cdot )\) and effect scale
\(h( \cdot )\) are identity functions (i.e., the estimand concerns
differences in mean outcomes), the marginal and conditional effect at
the average covariate values coincide, and direct transportability is
feasible under SEMA (Section 6.2). The mathematical intricacies of
marginal effects, which involve translating estimates across multiple
different scales, via \(h( \cdot )\) and \(g( \cdot )\), can
complicate transportability. In practice, it may be comparatively easier
to achieve transportability of conditional effects in indirect
comparisons, particularly when effect modification and effect
measurement are expressed on compatible scales (Section 6.3).

\hypertarget{collapsibility-and-its-role-in-transportability}{%
\subsection{Collapsibility and Its Role in
Transportability}\label{collapsibility-and-its-role-in-transportability}}

Collapsibility describes how statistical measures behave when averaging
across subgroups---particularly, whether a measure computed in subgroups
can be ``collapsed'' to give the same result in the overall
population.\textsuperscript{35,62,63} More precisely, collapsibility
refers to the property whereby marginal effects can be expressed as a
weighted average of conditional effects.\textsuperscript{35,62,63}
Direct collapsibility is a special case in which the population-average
conditional effect and marginal effect are mathematically equivalent
(i.e., the marginal effect coincides with an average of conditional
effects using population-representative weights).\textsuperscript{35,63}
For example, risk differences are directly collapsible, whereas (log)
risk ratios are collapsible but generally not directly collapsible.
Non-collapsibility refers to measures that are not
collapsible.\textsuperscript{35,62,63} Common examples include odds
ratios and hazard ratios. Table 1 summarizes the collapsibility
properties of the effect measures considered in this paper.

In the causal inference literature, collapsibility is conventionally
discussed as an intrinsic property of an effect measure for a direct,
within-population comparison (e.g., \(\Delta_{BA}\)). In anchored ITCs,
however, the estimand of interest is an anchored active--active contrast
(\(\Delta_{BC} = \Delta_{BA} - \Delta_{CA}\)), constructed indirectly
via a common comparator and therefore structurally different from such
within-population comparisons. As a result, the collapsibility
properties relevant for transportability depend not only on the effect
measure, but also on how the anchored contrast is constructed. If the
individual contrasts \(\Delta_{BA}\) and \(\Delta_{CA}\) are directly
collapsible, then \(\Delta_{BC} = \Delta_{BA} - \Delta_{CA}\) is also
directly collapsible. More subtly, even when the underlying
\(\Delta_{BA}\) and \(\Delta_{CA}\) contrasts are collapsible but not
directly collapsible, \(\Delta_{BC}\) exhibits direct collapsibility as
an induced property of the anchored contrast under SEMA and scale
alignment, a property that is not apparent from the collapsibility
properties of the individual contrasts alone. In this setting,
effect-modification terms cancel in the active--active contrast,
yielding marginal--conditional equivalence for \(\Delta_{BC}\), even
though the individual contrasts against the common comparator,
\(\Delta_{BA}\) and \(\Delta_{CA}\), may remain population-dependent
with distinct marginal and conditional effects. A formal proof of these
results is provided in Appendix B.

Transportability of marginal effects depends on whether the chosen
measure is collapsible and on what aspects of the covariate distribution
it depends upon. A recent simulation study by Remiro-Azócar empirically
demonstrates that both effect modifiers and purely prognostic factors
may modify marginal effects for measures that are not directly
collapsible.\textsuperscript{35} In other words, a prognostic factor on
the conditional scale may be a modifier of the marginal effect measure.
Thus, successfully transporting these types of marginal effects requires
adjusting for the entire joint distribution of purely prognostic factors
and treatment effect modifiers, which is often challenging to fully
address due to the limited data available for comparator studies (e.g.,
unknown correlations in comparator studies). These findings highlight a
critical distinction in transportability: whether an effect measure for
marginal effects depends on (i) marginal covariate means (or more
generally, moments) of effect modifiers or (ii) the full joint covariate
distribution of effect modifiers and prognostic factors. Non-collapsible
measures generally depend on the full joint covariate distribution,
including prognostic factors, making them more difficult to reliably
transport across study populations. Translated to Step 2 in pairwise
PAICs, recent work on NMAs suggests that direct transport of
\(\Delta_{BC(AC)}\) for a non-collapsible marginal effect would not be
biased by imbalances in purely prognostic factors only under restrictive
conditions: (i) the index and comparator studies are each homogeneous
with respect to those prognostic factors or (ii) both studies have
identical distributions of those prognostic factors.\textsuperscript{64}
However, in practice, distributions of prognostic factors typically vary
within and across studies, necessitating adjustment in transportability
analyses.

Furthermore, Remiro-Azócar concludes that directly collapsible effect
measures \emph{``facilitate the transportability of marginal effects
between studies,''} as they only depend on marginal covariate
moments---typically the covariate means---of effect modifiers but not
prognostic factors.\textsuperscript{35} In this paper we present
illustrative examples to further clarify that directly collapsible effect
measures depend on marginal covariate moments if and only if effect
modifiers and the effect measure both operate on the linear predictor
scale. However, if the scales differ, even directly collapsible
measures---such as the risk difference when derived from a logit
model---generally depend on the full joint covariate distribution,
including both effect modifiers and purely prognostic factors (Section
6.3).

Collapsibility and transportability are sometimes conflated. The
distinction between collapsibility as a within-population property and
transportability as a cross-population property of an estimand is
discussed in more detail in Appendix D.

\hypertarget{scale-of-effect-modification-and-correspondence-with-effect-measure}{%
\subsection{Scale of Effect Modification and Correspondence with Effect
Measure}\label{scale-of-effect-modification-and-correspondence-with-effect-measure}}

When considering the transportability properties of an effect, it is
crucial to understand the relationship between the scale of effect
modification (e.g., additive or multiplicative) and the effect measure
(e.g., risk difference, risk ratio, odds ratio). Whether an effect
\(\Delta_{BC}\) is directly transportable depends on the scale on which
effect modification is modeled and how it corresponds to the chosen
effect measure. If these scales do not align, both marginal and
population-average conditional treatment effects generally depend on the
covariate distribution, preventing direct transportability---even when
SEMA holds.

To formalize this, we distinguish between two scales: the model (linear predictor) scale, defined by the link function \(g(\cdot)\), and the effect measure scale, defined by the transformation \(h(\cdot)\). Here, the linear predictor scale refers to a modeling framework in which the effects of treatment, prognostic factors, and effect modifiers are assumed to be additive after application of the model link; this does not require linearity in the original covariates themselves. The mapping \(h \circ g^{-1}\) therefore determines the relationship between the model and effect measure scales. When \(h \circ g^{-1} = I\), the scales are aligned, which may facilitate transportability, but in general, these two scales need not coincide. For example, a logistic model targeting a log odds ratio has \(g(\cdot) = h(\cdot) = \operatorname{logit}(\cdot)\) and therefore the scales align as \(h \circ g^{-1} = I\), whereas the same model targeting a risk difference has \(g(\cdot) = \operatorname{logit}(\cdot)\) and \(h(\cdot) = I(\cdot)\), so \(h \circ g^{-1} = \operatorname{expit} \neq I\) and the scales are not aligned. The implications of scale alignment are formalized below.

Under this representation, the outcome model can be written as

\begin{equation}
\begin{aligned}
g\!\left(
\mathbb{E}\{Y \mid T=t, X\}
\right)
&=
\eta_t(X)
\\
&=
\beta_0
+
Z(X)^{T}\beta^{\mathrm{PF}}
+
\left(
Z(X)^{T}\beta_t^{\mathrm{EM}}
+
\delta_t
\right)
\mathbf{1}(t \neq A).
\end{aligned}
\label{eq:model_eta}
\end{equation}

where \(\eta_{t}(X)\) denotes the linear predictor under treatment
\(t\), $\beta^{\mathrm{PF}}$ are prognostic effects, \(\delta_{t}\) are
treatment main effects, and $\beta_t^{\mathrm{EM}}$ represent effect
modification. Here, \(Z(X)\) denotes the vector of covariates included
in the linear predictor, which may involve transformed terms such as
polynomials, logs, or interactions. Treatment effects are then defined as contrasts between model-based predictions after applying the transformation \(h(\cdot)\), generally expressed as \(\psi(a,b) = h(b) - h(a)\). As in STC, this model assumes a common intercept across studies.

Combining Equations (2) and (13), a general expression for the
population-average conditional effect is:

\begin{equation}
\Delta_{BA}^{\mathrm{Cond}}
=
\mathbb{E}_{X}\!\left\{
h\!\left(
g^{-1}\!\left(\eta_{B}(X)\right)
\right)
-
h\!\left(
g^{-1}\!\left(\eta_{A}(X)\right)
\right)
\right\}.
\label{eq:cond_effect_general}
\end{equation}

This broader formulation allows for the estimation of population-average conditional effects even when the model is fit on one scale (linear predictor via \(g(\cdot)\)), but the effect measure of interest is defined on a different scale (effect measure scale via \(h(\cdot)\)). Under scale alignment, \(h \circ g^{-1} = I\), and \(\Delta_{BA}^{\mathrm{Cond}}\) therefore depends only on the expected value of \(Z(X)\) rather than the full covariate distribution. If \(Z( \cdot )\) includes
polynomial functions of \(X\), then $\Delta_{BA}^{\mathrm{Cond}}$ can
equivalently be expressed in terms of the corresponding moments of
\(X\).

\begin{equation}
\begin{aligned}
\Delta_{BA}^{\mathrm{Cond}}
&=
\mathbb{E}_{X}\!\left\{
I\!\left(\eta_{B}(X)\right)
-
I\!\left(\eta_{A}(X)\right)
\right\}
\\
&=
\mathbb{E}_{X}\!\left\{
Z(X)^{\top}\beta_{B}^{\mathrm{EM}}
+
\delta_{B}
\right\}
\\
&=
\mathbb{E}\!\left\{Z(X)\right\}^{\top}
\beta_{B}^{\mathrm{EM}}
+
\delta_{B}.
\end{aligned}
\label{eq:cond_identity}
\end{equation}

Further, when the model
contains only linear terms in the original covariates (i.e.,
\(Z(X) = X\)), the population-average conditional effect depends only on
the mean of the covariates \(X\). In this case, Equation (15) simplifies
and provides a model-based representation of Equation (3):

\begin{equation}
\Delta_{BA}^{\mathrm{Cond}}
=
\bar{x}^{\top}\beta_{B}^{\mathrm{EM}}
+
\delta_{B}
=
\eta_{B}(\bar{x})
-
\eta_{A}(\bar{x}).
\label{eq:cond_mean_x}
\end{equation}

The corresponding marginal effect is:

\begin{equation}
\Delta_{BA}^{\mathrm{Marg}}
=
h\!\left(
\mathbb{E}_{X}\!\left\{
g^{-1}\!\left(\eta_{B}(X)\right)
\right\}
\right)
-
h\!\left(
\mathbb{E}_{X}\!\left\{
g^{-1}\!\left(\eta_{A}(X)\right)
\right\}
\right).
\label{eq:marg_effect_general}
\end{equation}

The key distinction is the ordering of transformation and averaging. For
conditional effects, the transformation \(h( \cdot )\) is applied before
averaging over the covariate distribution; for marginal effects,
averaging occurs first and the transformation is applied afterward.

Marginal and conditional effects generally differ due to nonlinearity of
the model link and/or effect measure scale. However, they coincide when
both the link and effect measure scale are linear (i.e.,
\(g( \cdot ) = h( \cdot ) = I( \cdot )\)), or in degenerate cases (e.g.,
no heterogeneity). Under the special case of linearity with
\(g( \cdot ) = h( \cdot ) = I( \cdot )\), averaging and transformation
commute:

\begin{equation}
h\!\left(
\mathbb{E}_{X}\!\left\{
g^{-1}\!\left(\eta_{t}(X)\right)
\right\}
\right)
=
\mathbb{E}_{X}\!\left\{
\eta_{t}(X)
\right\}.
\label{eq:h_g_inverse_identity}
\end{equation}

so, $\Delta_{BA}^{\mathrm{Marg}}=\Delta_{BA}^{\mathrm{Cond}}$, as in Equations (15) --
(16), implying that the marginal effect depends only on the expected
value of \(Z(X)\) rather than its full distribution. In contrast, when
\(g( \cdot )\) or \(h( \cdot )\) is nonlinear, then expectation and
transformation do not commute, and $\Delta_{BA}^{\mathrm{Marg}}$ generally
depends on the full covariate distribution, further complicating
transportability efforts.

The discussion here focuses on a single contrast against the common
comparator. As shown in Section 6.4, anchored active-to-active contrasts
\(\Delta_{BC}\) may have different transportability properties under
additional assumptions, such as SEMA. An effect may be directly
transportable only if both the effect modification and the effect
measure operate on the linear predictor scale (i.e.,
$h \circ g^{-1} = I$) for the contrast being transported;
importantly, this condition is necessary but not sufficient. Direct
transportability also requires assumptions ensuring that the resulting
contrast is invariant to the covariate distribution. How this invariance
arises for anchored active-to-active contrasts---through SEMA,
collapsibility, and scale alignment---and under what conditions it holds
is summarized in Section 6.4.

\hypertarget{conditions-for-direct-transportability}{%
\subsection{Conditions for Direct
Transportability}\label{conditions-for-direct-transportability}}

We illustrate in Section 7 that marginal indirect effects
\(\Delta_{BC}^{\text{Marg}}\) are directly transportable under the
following joint conditions: (i) SEMA holds; (ii) both the effect
modifiers and the effect measure operate on the same linear predictor
scale; and (iii) the effect measure is collapsible. When these
conditions are jointly satisfied, the resulting marginal estimand
\(\Delta_{BC}\) is induced to exhibit marginal--conditional equivalence
(i.e., marginal and conditional effects coincide, analogous to the
concept of direct collapsibility) and is directly transportable, meaning
it is invariant to changes in the covariate distribution across
populations. In contrast, population-average conditional effects are
directly transportable if SEMA is met and effect modifiers and the
effect measure both operate on the linear predictor scale---regardless
of whether the measure is collapsible. In both cases, direct
transportability reflects a property of the anchored estimand under
structural assumptions, rather than an intrinsic feature of PAIC methods
or effect measures. Notably, conditional linearity of the treatment
effect in the covariates is not required for direct transportability.

To illustrate the fundamental role of SEMA in this result, consider the
population-average conditional effects for \(B\) vs. \(A\) and \(C\) vs.
\(A\) under scale alignment (Equation 15) and shared effect
modification, where effects are defined on the linear predictor scale:

\begin{equation}
\begin{aligned}
\Delta_{BA}^{\mathrm{Cond}}
&=
\mathbb{E}\!\left\{Z(X)\right\}^{\top}\beta^{\mathrm{EM}}
+
\delta_{B},
\\
\Delta_{CA}^{\mathrm{Cond}}
&=
\mathbb{E}\!\left\{Z(X)\right\}^{\top}\beta^{\mathrm{EM}}
+
\delta_{C}.
\end{aligned}
\label{eq:cond_contrasts}
\end{equation}

where $\beta_{B}^{\mathrm{EM}}=\beta_{C}^{\mathrm{EM}}
\equiv \beta^{\mathrm{EM}}$ represents
shared effect modification. The anchored active--active contrast is then

\begin{equation}
\Delta_{BC}^{\mathrm{Cond}}
=
\Delta_{BA}^{\mathrm{Cond}}
-
\Delta_{CA}^{\mathrm{Cond}}
=
\delta_{B}
-
\delta_{C}.
\label{eq:cond_active_active}
\end{equation}

which no longer depends on \(X\). Thus, under SEMA, effect-modification
terms cancel in the active--active comparison, yielding a conditional
effect that is invariant to the covariate distribution and therefore
directly transportable. When combined with collapsibility for marginal
effects, this invariance leads to marginal--conditional equivalence for
\(\Delta_{BC}\) (Section 6.2). Under these conditions, the marginal
estimand of a collapsible measure can be written as a weighted average
of individual-level conditional contrasts (Appendix B, Proposition B1),

\begin{equation}
\Delta_{BC}^{\mathrm{Marg}}
=
\int
\Delta_{BC}^{\mathrm{Cond}}(x)\,
w(x)\,
dx.
\label{eq:marg_from_cond}
\end{equation}

where the weights are constrained to be positive and
$\int_{\mathcal{X}} w(x)\,dx = 1$.

By SEMA and scale alignment,
\(\Delta_{BC}^{\text{Cond}}(x) = \delta_{B} - \delta_{C}\), so

\begin{equation}
\Delta_{BC}^{\mathrm{Marg}}
=
\left(
\delta_{B}
-
\delta_{C}
\right)
\int w(x)\,dx
=
\delta_{B}
-
\delta_{C}.
\label{eq:marg_active_active}
\end{equation}

which is invariant to the covariate distribution \(P\) and therefore
directly transportable. In contrast, for non-collapsible measures, such
a weighted-average representation generally does not hold, and marginal
effects depend on the full covariate distribution (Section 6.3).

Figure 3 summarizes these results in a high-level framework for determining when unadjusted indirect comparisons are sufficient and when population adjustment is required to address differences in effect modifiers or prognostic factors across populations.
The formal conditions under which conditional and marginal
active-to-active contrasts are directly transportable are stated as
Propositions A1--A2 and proved in Appendix A. A more comprehensive
summary of the requirements for direct transportability is provided in
Section 8. Figure 4 further synthesizes these considerations into a decision
framework showing how estimand choice (conditional vs marginal),
collapsibility of the effect measure, scale alignment, and SEMA jointly
determine whether PAIC-derived treatment effects are directly
transportable or population-specific. Note that direct transportability
here pertains to the active-to-active contrast \(\Delta_{BC}\) and does
not imply that the underlying individual contrasts from each trial
(\(\Delta_{BA}\) and \(\Delta_{CA}\)) are also invariant across
populations.

\vspace{5mm}
\noindent\textbf{Figure 3.} Decision Framework for Determining When Unadjusted NMAs
Identify a Transportable Treatment Effect Estimand and When PAICs Are
Required

\includegraphics[scale=0.63]{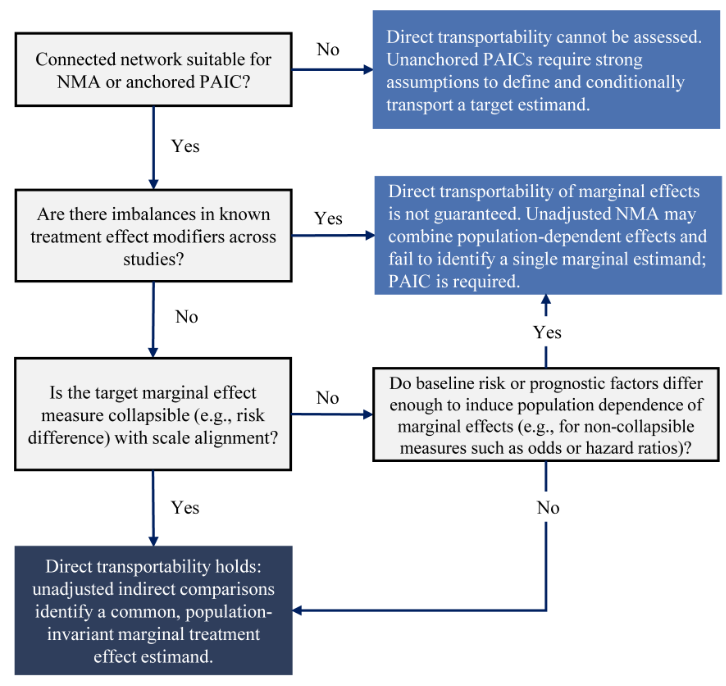}
\vspace{2mm}

{\footnotesize
\setlength{\baselineskip}{0.9\baselineskip}
\noindent Figure 3 focuses on direct transportability of unadjusted indirect
comparisons of \(\Delta_{BC}\); population invariance of marginal
estimands may arise either from the absence of effect modification or,
in special cases, from structural properties of the estimand (e.g.,
collapsibility on a compatible scale). When these conditions are not
met, the issue is not merely failed transportability but the absence of
a single marginal estimand shared across studies, and unadjusted NMAs
combine population-dependent effects rather than identifying a common
marginal estimand.\par
}

\vspace{5mm}

\noindent\textbf{Figure 4.} Framework for Selecting PAIC Methods and Assessing
Transportability of Treatment Effects Under Estimand Choice,
Collapsibility, and SEMA

\includegraphics[scale=0.84]{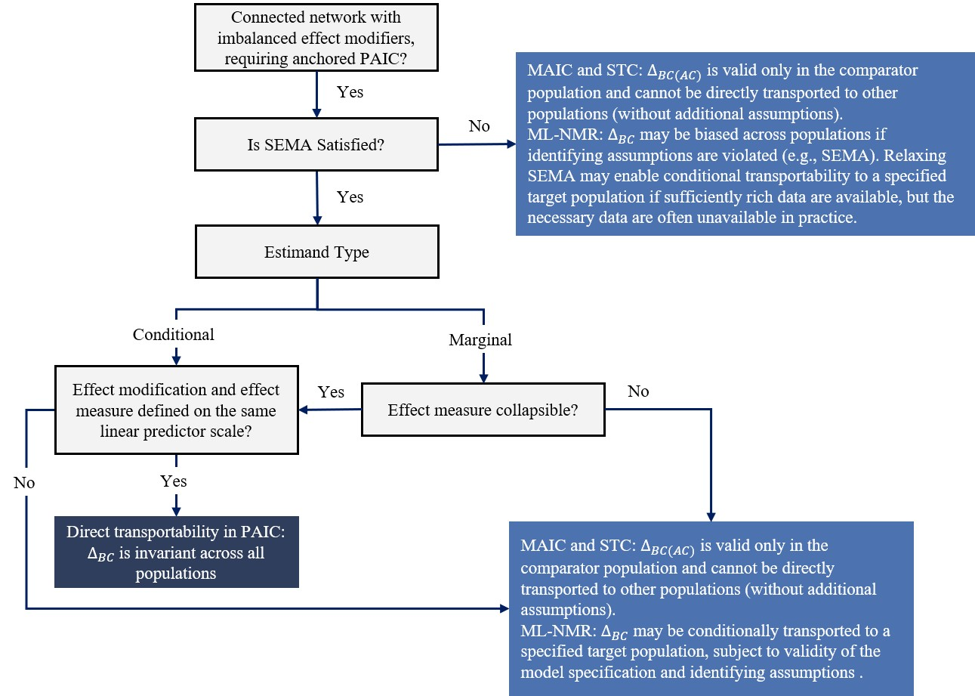}

\vspace{2mm}

{\footnotesize
\setlength{\baselineskip}{0.9\baselineskip}
\noindent Even when the SEMA holds, direct transportability of \(\Delta_{BC}\)
across populations depends on the target estimand (conditional vs
marginal), the collapsibility of the effect measure, and alignment
between the scale of effect modification and the effect measure. For
commonly used non-collapsible marginal measures (e.g., hazard ratios),
MAIC and STC estimates are generally valid only in the comparator
population, provided the underlying assumptions are met. ML-NMR may support conditional transport to relevant target
populations, but this relies on additional modeling assumptions and does
not, in general, guarantee direct transportability.\par
}

\hypertarget{illustrative-examples-of-two-step-transportability-in-maicsstcs}{%
\section{Illustrative Examples of Two-step Transportability in
MAICs/STCs}\label{illustrative-examples-of-two-step-transportability-in-maicsstcs}}

In this section, we use illustrative numerical examples to examine the direct transportability of active-to-active indirect comparisons across populations, corresponding to Step 2 of the two-step transport process described in Section 4. Prior simulation studies have primarily examined the
statistical performance of MAICs and STCs for Step 1 under model
misspecification, limited overlap, and finite-sample
constraints.\textsuperscript{34,35,48,65-70} Here,
however, we deliberately isolate Step 2, which is often overlooked in
practice. We treat Step 1 as successful by assuming that 
\(\Delta_{BC(AC)}\) is estimated without bias in the comparator population and examine whether this estimate remains invariant when transported to populations with different covariate distributions.

We illustrate these transportability properties across five different
effect measures:

\begin{enumerate}
\def\labelenumi{\arabic{enumi}.}
\item
  Mean difference, estimated via linear regression
\item
  Log odds ratio, estimated via logistic regression
\item
  Log risk ratio, estimated via log-binomial regression
\item
  Difference in restricted mean survival time (RMST), estimated using
  Weibull regression
\item
  Log ratio of RMST, also estimated using Weibull regression
\end{enumerate}

For each analysis, we specified a regression model characterizing the
true relative effects of \(B\) vs. \(A\) and \(C\) vs. \(A\), adjusting
for a single covariate \(X\) that is both an effect modifier and
prognostic factor (i.e., the true outcome model). \(X\) was uniformly
distributed with mean \(\mu_{X(P)}\) (which is specific to a given
population P) and a fixed range of 2:
\(X\sim \mathrm{Unif}(\mu_{X(P)},\ \mathrm{range} = 2)\). This corresponds to the type of
outcome regression model commonly estimated in STC and ML-NMR. \(X\) was
centered at the mean of the comparator population. We simulate 21
populations with \(\mu_{X(P)}\) values ranging from -0.5 to 0.5 in
increments of 0.05, with each population comprising one million
individuals. Note that, in all illustrative examples, simulation was
used only to approximate the specified covariate distributions and
compute population-level estimands via Monte Carlo integration.

To investigate the transportability properties of each effect measure,
we derive marginal and population-average conditional effects for \(B\)
vs. \(C\) in each simulated population. The relative effect in the
comparator population (where \(\mu_{X} = 0\)) represents the adjusted
result that would be estimated via MAIC or STC under the conditional
transportability assumption. Equality of the effect estimate across all
21 populations would indicate direct transportability of \(\Delta_{BC}\)
(i.e., Step 2 in MAICs and STCs).

\hypertarget{mean-difference}{%
\subsection{Mean Difference}\label{mean-difference}}

The true outcome model for a continuous outcome
\(Y|X\sim Norm(\eta_{k(P)},\sigma^{2})\) with the identity link function
(i.e.,
\(g\left( \mathbb{E}\left\{ Y \middle| X \right\} \right) = \mathbb{E}\left\{ Y \middle| X \right\}\))
was defined as

\begin{equation}
\mathbb{E}\!\left\{Y \mid X\right\}
=
\eta_{k(P)}
=
\beta_{0}
+
x_{ik(P)}\beta_{1}
+
x_{ik(P)}\beta_{2,k}
+
\delta_{k}.
\label{eq:model_example}
\end{equation}

where \(k \in \{ A,B,C\}\) represents treatment group,
\(\beta_{0} = 20\) is the intercept (common baseline risk across all
populations), \(\beta_{1} = 10\) is the prognostic effect of \(X\),
\(\beta_{2,k}\) represents an interaction term for effect modification
(with \(\beta_{2,A} = 0\) by definition), and \(\delta_{k}\) is the
baseline treatment effect (vs. $A$), with \(\delta_{B} = 10\) and
\(\delta_{C} = 5\). We consider two scenarios regarding effect
modification. Under the SEMA, effect modification is assumed to be
shared across both active treatments, such that
\(\beta_{2,B} = \beta_{2,C} = 2\). In contrast, under the no-SEMA
scenario, effect modification is treatment-specific, with
\(\beta_{2,B} = 2\) and \(\beta_{2,C} = - 4\).

Figure 5 displays the bias incurred when directly transporting marginal
and population-average mean differences for \(B\) vs. \(C\) from the
comparator population (identified in Step 1 of an MAIC/STC) to target
populations with differing covariate distributions (Step 2). In the
linear setting, population-average conditional and marginal mean
differences are mathematically equivalent (i.e., direct collapsibility),
and thus are represented by the same line.

The vertical line at \(\mu_{X(AC)} = 0\) denotes the comparator
population, in which \(\Delta_{BC(AC)}\) is identified via MAIC or STC.
When SEMA holds, \(\Delta_{BC}\) is invariant across all populations,
resulting in zero transport bias across all values of \(\mu_{X(P)}\);
that is, the comparator-population estimate \(\Delta_{BC(AC)}\) is
unaffected by shifts in the distribution of \(X\) and can be directly
transported to any external target population under SEMA. However, in
the absence of SEMA, the bias increases linearly as the mean of \(X\)
diverges from the comparator population, implying that \(\Delta_{BC}\)
is not directly transportable across populations.

To illustrate this more concretely, we examine two hypothetical target
populations (of the 21 simulated to draw the figures): \(P_{2}\) with
\(\mu_{X} = - 0.1\) and \(P_{3}\) with \(\mu_{X} = - 0.4\). We compare
the adjusted effect in the comparator population (central reference
point, \(P_{1}\)) with the true effect in \(P_{2}\) and \(P_{3}\) (i.e.,
flagged points along the red line). Under the no-SEMA scenario, the
divergence between the adjusted effect in the comparator population and
true effect in the target grows as the mean of \(X\) shifts further from
zero (i.e., the true effect in \(P_{2}\) is closer to the comparator
population than \(P_{3}\)). In general, the magnitude of this
discrepancy depends on both the strength of the effect modification for
\(B\) vs. \(C\), calculated as \(\beta_{2,B} - \beta_{2,C}\), and the
extent of overlap in joint covariate distributions between populations.

\vspace{5mm}
\noindent\textbf{Figure 5.} Bias in Conditional and Marginal Mean Differences Under Direct
Transportability

\noindent \includegraphics[width=1\linewidth]{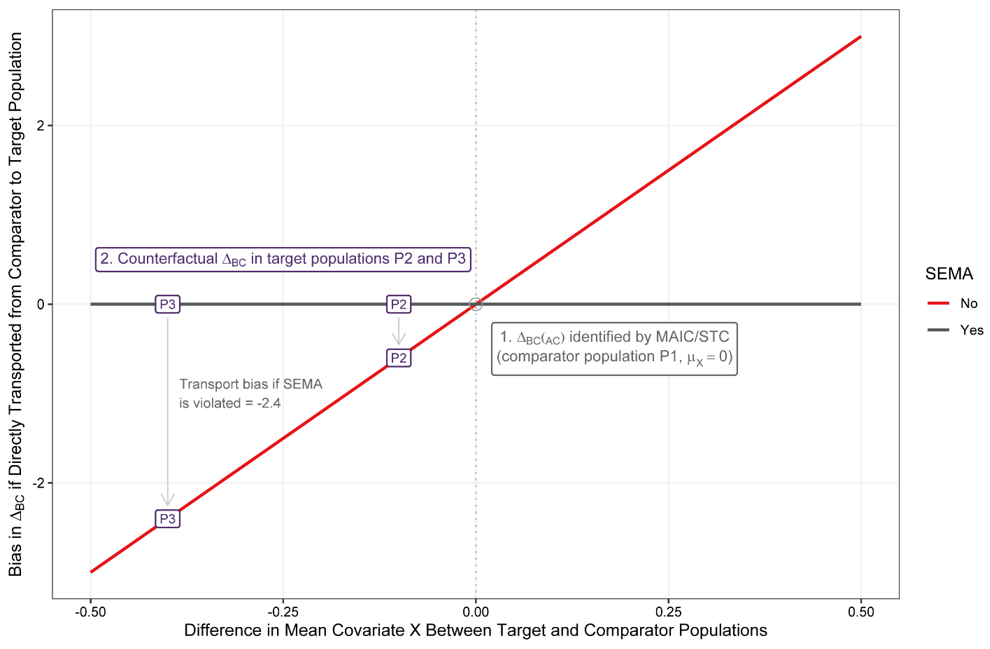}

{\footnotesize
\setlength{\baselineskip}{0.9\baselineskip}
\noindent Bias is shown as a function of the difference in the mean of covariate
\(X\) between the target and comparator populations. The vertical dashed
line denotes the comparator population (\(\mu_{X(AC)} = 0\)), in which
the PAIC estimand \(\Delta_{BC(AC)}\) is identified. A flat line
indicates direct transportability (i.e., zero Step-2 transport bias), as
is observed when SEMA holds. Any line or curve that is not completely
flat indicates that bias is induced in Step 2 by directly transporting
the effect derived in the comparator population to other target
populations.\par}

\hypertarget{log-odds-ratio-and-risk-ratio}{%
\subsection{Log Odds Ratio and Risk
Ratio}\label{log-odds-ratio-and-risk-ratio}}

Next, we consider a logistic regression model for a binary outcome. The
general model structure is identical to that in the previous section,
but the link function is logit, and the conditional distribution of the
outcome is \(Y|X\sim \mathrm{Binomial}\left( \mathrm{logit}^{- 1}(\eta_{k(P)}) \right)\).
We assume parameters \(\beta_{0} = 0\), \(\beta_{1} = - 1\),
\(\delta_{B} = - 3\), \(\delta_{C} = - 4\). Under the SEMA, we set
\(\beta_{2,B} = \beta_{2,C} = - 3\), while under the alternative
scenario without SEMA, \(\beta_{2,B} = - 3\) and \(\beta_{2,C} = - 4\).

Figure 6 displays the bias incurred when directly transporting
conditional and marginal log odds ratios for \(B\) vs. \(C\) from the
comparator population to target populations with differing covariate
distributions. For the conditional log odds ratio (left panel), the
general pattern mirrors that observed for mean differences: when SEMA
holds, the conditional log odds ratio for \(B\) vs. \(C\) is invariant
across populations, resulting in zero transport bias. When SEMA is
violated, transport bias increases linearly as the mean of $X$ diverges
from that of the comparator population, reflecting treatment-specific
effect modification.

In contrast, for the marginal log odds ratio (right panel), transport
bias occurs and varies with the covariate distribution even when SEMA
holds. This reflects the non-collapsibility property of the (log) odds
ratio. As marginal log odds ratios average over the covariate
distribution, they depend nonlinearly on both the conditional outcome
model and the joint distribution of covariates. As a result, marginal
log odds ratios are not directly transportable, regardless of whether
SEMA is satisfied. This example highlights a key distinction between
conditional and marginal estimands: while conditional effects defined on
the model's linear predictor scale may be directly transportable under
SEMA, marginal effects for non-collapsible measures generally are not.

\vspace{5mm}
\noindent\textbf{Figure 6.} Bias in Conditional and Marginal Log Odds Ratios Under Direct
Transportability

\noindent \includegraphics[width=1\linewidth]{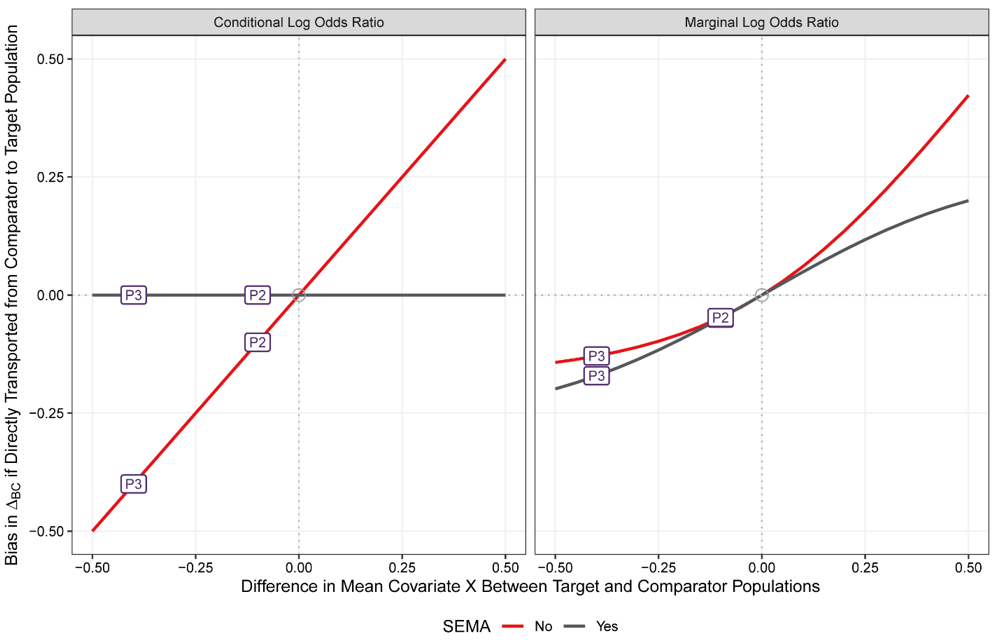}

{\footnotesize
\setlength{\baselineskip}{0.9\baselineskip}
\noindent Bias is shown as a function of the difference in the mean of covariate
\(X\) between the target and comparator populations. The vertical dashed
line denotes the comparator population, in which the PAIC estimand is
identified. Left panel: conditional log odds ratios exhibit zero
transport bias under SEMA, indicating direct transportability, but incur
bias when SEMA is violated. Right panel: marginal log odds ratios
exhibit population dependence even when SEMA holds, reflecting
non-collapsibility and failure of Step-2 direct transportability.\par}

\vspace{5mm}

Following the log odds ratio example, we use a log-link model to
generate a binary outcome with the same covariate structure and general
pattern of effect modification, but targeting the log risk ratio rather
than the log odds ratio (Appendix C). For the data-generating mechanism,
effect modification and SEMA are specified on the log-risk scale. Under
this log-risk model, the marginal log risk ratio for \(B\) vs. \(C\) is
invariant across populations under SEMA, mirroring the mean-difference
example (Figure 5). This contrasts with the log odds ratio example,
where non-collapsibility induces population dependence even when SEMA
holds. This illustrates that despite both being multiplicative measures,
risk ratios and odds ratios have fundamentally different
transportability properties. The log risk ratio result reflects the
contrast-induced direct collapsibility described in Section 6.2, under
which the anchored \(B\) vs. \(C\) log risk ratio is invariant across
populations when SEMA holds on the same log-link scale used to define
the effect measure.

\hypertarget{rmst-differences-and-ratios}{%
\subsection{RMST Differences and
Ratios}\label{rmst-differences-and-ratios}}

Although hazard ratios are commonly used for survival outcomes, their
interpretation and transportability rely on additional modeling
assumptions (e.g., proportional hazards must hold for comparisons across
all studies), and even when such assumptions hold, marginal hazard
ratios are inherently population-dependent due to non-collapsibility;
analogous to the patterns for the odds ratio shown in Figure 6,
conditional hazard ratios for \(\Delta_{BC}\) are invariant across
populations under SEMA, whereas marginal hazard ratios are not (Table
1), and may also vary over time. RMST-based measures are often advocated
as alternatives to hazard ratios, as they are collapsible and clinically
interpretable.\textsuperscript{71}

In the final scenario, we examine the properties of RMST-based measures
using a parametric Weibull survival model. The survival function for treatment arm \(k\) in population \(P\) is defined as
\begingroup
\abovedisplayskip=18pt
\belowdisplayskip=18pt
\begin{equation}
S_{k(P)}\left( t \middle| x \right)
=
\exp\!\left(
- t^{\nu}
\exp\!\left( \eta_{k(P)}(x) \right)
\right)
\end{equation}
\endgroup
where the shape parameter \(\nu = 1.5\) is assumed to be common across
populations, and \(\eta_{k(P)}\) follows the same general specification
as in Equation (23). The model includes a fixed intercept
\(\beta_{0} = - 1\) (i.e., common baseline risk), prognostic coefficient
\(\beta_{1} = \log(0.25)\), and baseline treatment effects
\(\delta_{B} = \log(0.4)\) and \(\delta_{C} = \log(0.6)\). Under SEMA we
assume \(\beta_{2,B} = \beta_{2,C} = \log(0.9)\); otherwise,
\(\beta_{2,B} = \log(0.7)\) and \(\beta_{2,C} = \log(0.9)\). As in
standard STC and ML-NMR implementations, effect modification is defined
on the log-hazard scale in this example. This example is adapted from
the artificial data-generating mechanism presented in Phillippo et al.
(2025).\textsuperscript{45}

Figure 7 displays the bias incurred when directly transporting
RMST-based contrasts from the comparator population to target
populations with differing covariate distributions. Because RMST
difference is a directly collapsible estimand, the population-average
conditional and marginal RMST differences coincide within each
population. However, despite being directly collapsible---analogous to
mean differences from linear regression---RMST differences are not
directly transportable under SEMA. Likewise, the conditional log ratio
of RMSTs is not directly transportable under SEMA. This finding may seem
counterintuitive, but it reflects the scale on which effect modification
is defined relative to the effect measure. In this case, SEMA and effect
modification are specified on the log hazard scale (via the model link
\(g( \cdot )\)), while RMST is defined from the model-implied survival curve (via
\(h( \cdot ))\), so \(h( \cdot ) \neq g( \cdot )\). The non-linear
transformation from hazards to RMST therefore introduces dependence on
the entire covariate distribution, preventing cancellation of
effect-modification terms under SEMA. As a result, direct
transportability fails due to scale misalignment rather than
non-collapsibility. For survival outcomes, \(h( \cdot )\) is interpreted more generally as a function of the model-implied survival curve, rather than as a pointwise transformation of a scalar mean outcome. For RMST, \(h( \cdot )\) maps the predicted survival curve to the area under that curve up to a specified time point.

\vspace{5mm}
\noindent\textbf{Figure 7.} Bias in RMST-based Conditional and Marginal Contrasts Under
Direct Transportability

\noindent \includegraphics[width=1\linewidth]{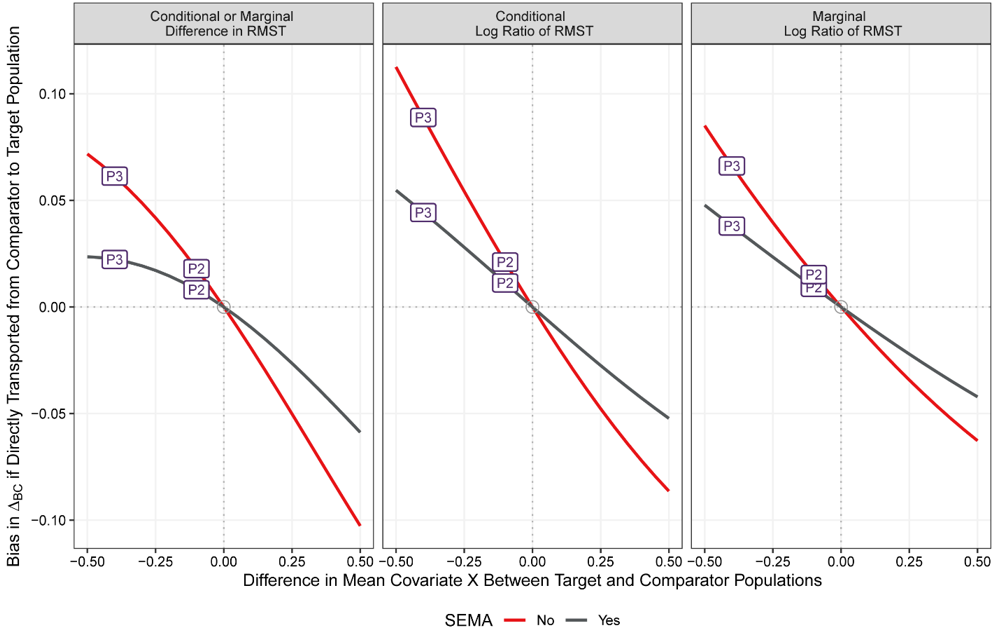}

{\footnotesize
\setlength{\baselineskip}{0.9\baselineskip}
\noindent Bias is shown as a function of the difference in the mean of covariate
\(X\) between the target and comparator populations. Left panel: RMST
difference; middle panel: conditional log ratio of RMST; right panel:
marginal log ratio of RMST. The vertical dashed line denotes the
comparator population in which the PAIC estimand is identified. Despite
RMST difference being a directly collapsible estimand, bias varies with
the covariate distribution even when SEMA holds. This pattern of
transport bias reflects scale misalignment between effect modification
(log-hazard scale) and the RMST estimand (time scale) rather than
non-collapsibility.\par}

\hypertarget{summary-of-requirements-for-direct-transportability}{%
\section{Summary of Requirements for Direct
Transportability}\label{summary-of-requirements-for-direct-transportability}}

Building on the illustrative examples presented in Section 7, Table 1
summarizes the mathematical properties of common effect measures in
PAICs, with a focus on the requirements for direct transportability of
\(\Delta_{BC}\). It presents 14 effect measures, including the five
effect measures featured in the earlier illustrative examples, and
synthesizes the decision frameworks in Figures 3-4 by mapping estimand
choice, collapsibility, scale alignment, and SEMA to the
transportability properties of these measures. The general pattern that
emerges across Table 1 shows that when effect modifiers and the effect
measure operate on the linear predictor scale and SEMA is justifiable,
\(\Delta_{BC}^{\mathrm{Cond}}\) is directly transportable across populations in
STCs and ML-NMR; this is because under these conditions the individual
contrasts \(\Delta_{BA(AC)}\) and \(\Delta_{CA(AC)}\) are typically a
function of the covariate moments (usually, covariate means) and their
effects cancel out in active-to-active indirect comparisons. By
contrast, marginal effects for \(B\) vs. \(C\) \(\Delta_{BC}^{\mathrm{Marg}}\)
are only directly transportable if (i) SEMA is justifiable; (ii) EMs and
the effect measure operate on the linear predictor scale at the
conditional level; and (iii) the effect measure is collapsible. Marginal
hazard ratios and odds ratios are the most common measures compared in
PAICs, but due to non-collapsibility are generally population-dependent
and not directly transportable (Table 1).

Although direct transportability may be desirable, particularly in MAIC
and STC, it should not necessarily dictate which effect measures are
targeted. Rather, the choice should be guided by the research
question(s) and plausibility of the underlying modeling assumptions. For
instance, comparing risk differences in an MAIC may facilitate direct
transportability of \(\Delta_{BC}^{\mathrm{Marg}}\) under SEMA, enabling
statistical inferences in the index population (as opposed to being
restricted to the comparator population). However, assuming SEMA holds
on the risk scale would typically be a much stronger assumption than
SEMA on the log or logit scales, and may not be clinically justifiable.
If SEMA is justifiable only on the logit scale but not the risk scale
and the risk difference is the effect measure of interest, then neither
the marginal nor the conditional risk difference for \(B\) vs. \(C\)
would be directly transportable.

For survival outcomes, ML-NMR and STC models specify effect modification
and SEMA on the log hazard scale, even when targeting an effect measure
besides the hazard ratio, such as the marginal RMST differences and
ratios. In this setting, effect modification on the RMST scale is not
modeled directly but is indirectly induced through the assumed hazard
model. Due to this scale misalignment, these measures are not directly
transportable under the standard hazard-scale SEMA specification.
Although effect modification could in theory be modeled directly on the
RMST scale, which may be more natural for certain applications, doing so
would require re-evaluating the plausibility of SEMA on that scale.
Additionally, while several regression- and weighting-based methods have
been proposed to estimate RMST differences using an identity link while
adjusting for baseline covariates\textsuperscript{72-74}, these are
treated as working models for estimation rather than structural models
that imply linearity of RMST in covariates. While direct
transportability of RMST-based effect measures could arise under strong
structural assumptions, such assumptions represent a substantive shift
from conventional survival modeling and may require strong clinical or
empirical justification.

Appendix D summarizes the population dependence of marginal
active--active contrasts for the effect measures in Table 1.

\vspace{4mm}

\noindent \textbf{Table 1.} Collapsibility and Transportability Properties of
\(\mathbf{\Delta}_{\mathbf{BC}}\) Under Different Effect Measures

\noindent \includegraphics[
    scale=.74,
    trim=0.1cm 11.4cm 0.1cm 0.1cm,
    clip
]{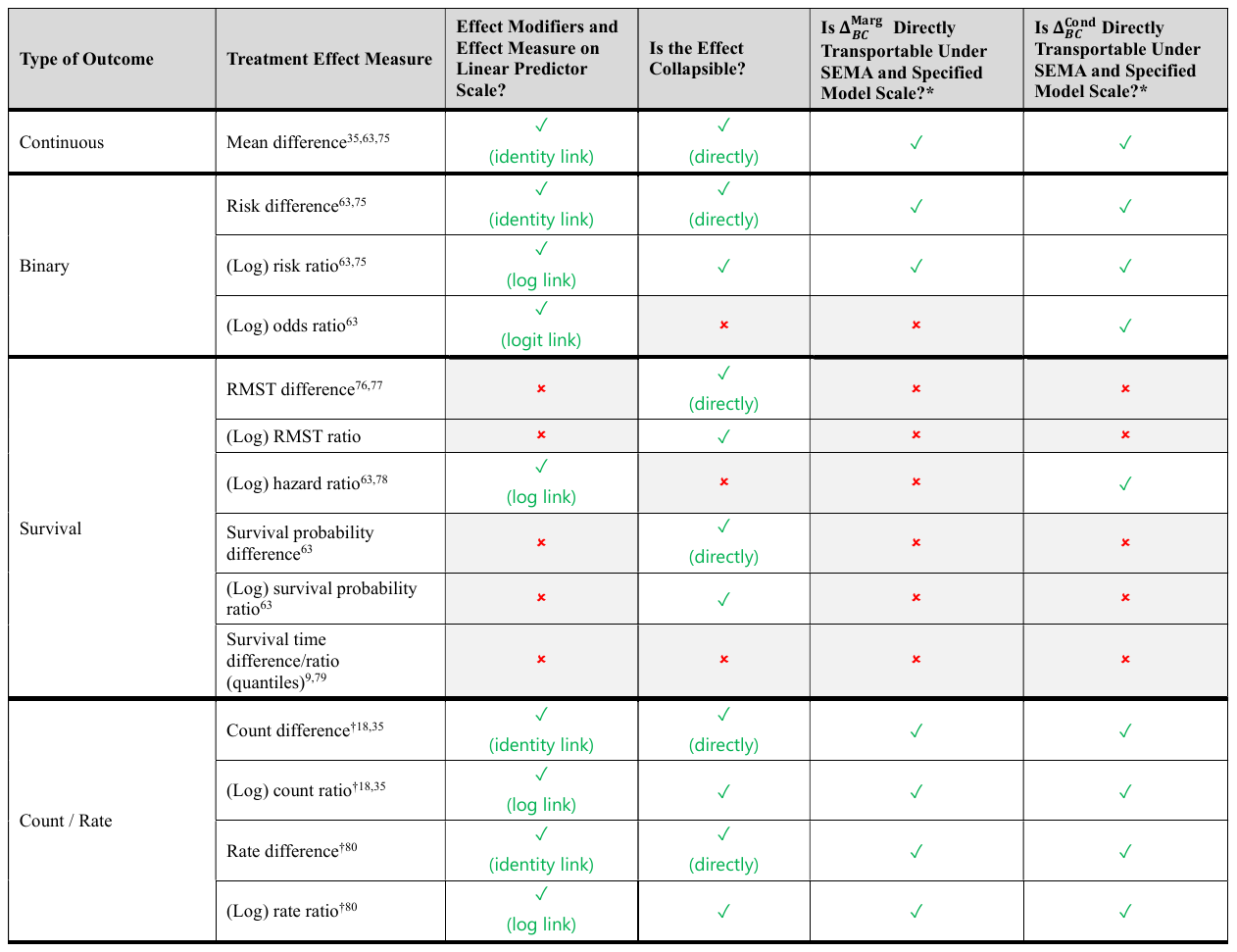}

{\footnotesize
\setlength{\baselineskip}{0.9\baselineskip}
\noindent *Direct transportability is evaluated under SEMA and the scale alignment specified for each effect measure. For conditional active-to-active contrasts, SEMA and scale alignment jointly ensure direct transportability; for marginal contrasts, collapsibility is additionally required. A checkmark indicates that direct transportability is guaranteed under the specified assumptions, whereas a cross indicates that it is not generally guaranteed. When direct transportability holds, \(\Delta_{BC}\) is invariant to changes in the covariate distribution across populations. Otherwise, explicit standardization to the target population may be required.

(Log) indicates either the ratio measure or its natural logarithm.
Because collapsibility and transportability are invariant to monotone
transformations, these are grouped. 

\textsuperscript{†}Counts are defined over a common exposure period, while rates are defined per unit exposure (e.g., via a regression offset). Conventional person-time rate differences and ratios are generally non-collapsible because exposure times may depend on treatment and prognostic factors.\textsuperscript{80} For PAICs, we define marginal rates by averaging predicted individual rates over a common target-population distribution, assuming a common exposure-time distribution across treatments; rate contrasts then inherit the corresponding collapsibility and transportability properties of count contrasts.\par}

\hypertarget{extensions-of-transportability-in-paics}{%
\section{Extensions of Transportability in
PAICs}\label{extensions-of-transportability-in-paics}}

\hypertarget{extension-from-pairwise-to-multi-arm-comparisons}{%
\subsection{Extension from Pairwise to Multi-arm
Comparisons}\label{extension-from-pairwise-to-multi-arm-comparisons}}

The preceding sections focus on transportability in the binary treatment
setting (\(B\) vs. \(A\)). These concepts of direct and conditional
transportability extend naturally to trials and evidence networks
involving more than two treatments, denoted by the
set\(\mathcal{\ T \in \{}A,\ B,\ C,\ D,\ \ldots\}\). In this setting,
each individual has a vector of potential outcomes
\{\(Y^{t}:t \in \mathcal{T}\)\}, one for each treatment in the set
\(\mathcal{T}\). The causal effect of interest may involve comparing any
pair of treatments in the set \(\mathcal{T}\). If there are \(k\)
treatments in set \(\mathcal{T}\), then there exists
\(\binom{k}{2} = \frac{k(k - 1)}{2}\) unique pairs of treatments that
could be compared on a pairwise basis.

In general, the transported effect from source population to target
population for a generic, unique pair \(t,s \in \mathcal{T}\) is then:

\begin{equation}
\Delta_{P_{1}\to P_{2}}^{t,s}
=
\psi\!\left(
  \mu_{P_{1}\to P_{2}}^{t},\,
  \mu_{P_{1}\to P_{2}}^{s}
\right)
\end{equation}

All considerations discussed earlier---estimand choice, collapsibility,
scale alignment, and transportability assumptions---apply analogously in
the multi-arm setting.

\hypertarget{extending-transportability-in-anchored-networks-ml-nmr}{%
\subsection{Extending Transportability with ML-NMR: Methodological Considerations}\label{extending-transportability-in-anchored-networks-ml-nmr}}

Methods such as MAIC and STC were originally developed for two-arm
settings and can conditionally transport effects only for treatments
observed in the index study. ML-NMR extends STC and NMA to connected
treatment networks by synthesizing information across all studies and
treatments simultaneously.\textsuperscript{81} Rather than relying on a
single source population, as in MAIC and STC, ML-NMR fits a hierarchical
outcome model that incorporates all study-specific effects, prognostic
factors, treatment indicators, and treatment--covariate interactions
(effect modifiers). In this sense, the evidence base supporting
transport in ML-NMR is a composite of all studies in the connected
network rather than a single index trial.

By leveraging information from the entire network, ML-NMR can estimate
relative effects for any connected treatment contrast and evaluate them
in pre-specified target populations. This distinguishes ML-NMR from
pairwise MAIC and STC, which naturally yield estimates specific to the
comparator population. However, this flexibility requires strong
identification assumptions and changes the role of SEMA. Importantly,
the two-step transport framework discussed in Section 4 does not
directly apply to ML-NMR. In pairwise MAIC and STC, the first transport
step---estimating \(\Delta_{BA(AC)}\), and subsequently
\(\Delta_{BC(AC)}\), in the comparator population---does not require
SEMA; rather, it requires correct adjustment for all effect modifiers of
\(B\) vs. \(A\). SEMA may become relevant if a second transport step is
invoked, where the resulting active-to-active contrast is assumed to be
directly transportable beyond the comparator population. In ML-NMR, by
contrast, SEMA is usually operationalized within the outcome model
itself through shared treatment--covariate interaction terms, so the
assumption enters during model estimation and carries through to
post-estimation steps, including transport. Although SEMA could in
principle be relaxed with sufficiently rich data to estimate
treatment-specific effect modification across the network, ML-NMR
usually relies on SEMA in practice due to data constraints.

This distinction has practical consequences if SEMA is violated.
Pairwise MAIC and STC estimates may be valid for the comparator
population, provided the Step 1 conditional transport assumptions are
satisfied. For ML-NMR, however, validity in the comparator population may not necessarily be preserved when SEMA is violated, even in a two-study setting; the implications of SEMA violation may depend on the outcome type and estimand. 
For binary outcomes, a simulation study by Phillippo et al. suggests that estimates for \(C\) vs. \(A\) in the
comparator population are robust to
misspecification of SEMA (i.e., remain unbiased), likely because
the arm-specific intercepts in ML-NMR can calibrate predictions to match
the observed aggregate event proportions.\textsuperscript{34} However,
recent simulations by Chandler and Ishak (2026) in the unanchored setting (using two-parameter Weibull survival models) suggest that this
robustness may not generally extend to comparisons of survival outcomes, because survival
models must characterize the survival time distribution over follow-up
rather than a single event proportion;\textsuperscript{82} consequently,
a simple baseline shift (i.e., intercept adjustment) may match survival at one time point but will not
generally recover the full survival curve, potentially biasing survival
probabilities and marginal estimands even in the comparator
population.\textsuperscript{82} More flexible baseline hazard models, including spline-based specifications, may mitigate this limitation by better reproducing marginal survival distributions despite misspecified conditional effects, although shared model constraints across treatment arms may still limit their ability to do so. Overall, the extent of potential bias
could depend on the available data for relaxing SEMA, the outcome type
and estimand, the flexibility of the baseline hazard model and constraints imposed across treatment arms, differences in relevant covariate distributions, and
the magnitude of the violation. Further research is needed to assess the
robustness of ML-NMR to violations of SEMA and other forms of model
misspecification, and to identify methodological extensions and data
requirements that could relax these identification
assumptions.\textsuperscript{83}

ML-NMR extends the outcome-regression logic of STC by embedding it
within an NMA framework. Thus, it handles baseline risk differently from
STC and MAIC. In STC, the outcome model is fitted in the index study and
includes a single intercept anchored to the index study's baseline risk.
Transport to \(P_{2}\) is achieved by averaging predictions over the
target covariate distribution. As a result, STC implicitly assumes that
any between-study differences in baseline risk are fully explained by
the observed covariates included in the model. MAIC makes a similar
assumption regarding baseline risk, but this assumption is invoked
implicitly through reweighting. This distinction is especially relevant
when absolute outcomes or marginal effects depend on baseline risk.

Conversely, ML-NMR allows for study-specific intercepts:

\begin{equation}
g\!\left(
\mathbb{E}\{Y \mid T=t, X\}
\right)
=
u_{s}
+
Z(X)^{\top}\beta^{\mathrm{PF}}
+
\left(
Z(X)^{\top}\beta_{t}^{\mathrm{EM}}
+
\delta_{t}
\right)
\mathbf{1}(t \neq A).
\label{eq:study_intercept_model}
\end{equation}

where \(u_{s}\) captures baseline risk in study \(s\). By including
\(u_{s}\), ML-NMR allows baseline risk to vary across studies even after
adjusting for observed covariates. This distinction becomes important
when prognostic factors are incompletely measured or reported, because
ML-NMR can accommodate residual heterogeneity in baseline risk more
naturally than MAIC or STC. This increased flexibility may require
additional modeling assumptions, as \(u_{s}\) must be specified or
estimated for each study population; when the target population is
external to the network, its baseline risk is not observed directly and
must be estimated from external data (e.g., calibrating an intercept to
reproduce the reported marginal outcome under the ML-NMR model fitted to
other studies) or linked to the study-specific intercepts in the
network, each requiring additional assumptions. Thus, while
study-specific intercepts may address one source of between-study
heterogeneity (i.e., baseline risk differences), they also introduce an
additional layer of uncertainty in PAICs.

These considerations apply to anchored networks, where ML-NMR can use
the common comparator structure to synthesize relative effects across
populations. In unanchored settings, however, the absence of a common
comparator makes it substantially harder to distinguish true treatment
differences from differences in study populations. In these cases,
relative effects must be inferred from outcomes observed in different
populations, placing greater reliance on assumptions about prognostic
factors, effect modification, and baseline risk.

\hypertarget{challenges-in-the-unanchored-setting-ml-umr}{%
\subsection{Challenges in the Unanchored Setting:
ML-UMR-SA}\label{challenges-in-the-unanchored-setting-ml-umr}}

Unanchored single-arm PAICs involve transporting outcomes for treatments from fully
disconnected studies across populations. In HTA,
unanchored comparisons of fully disconnected studies are far more common than anchored comparisons;
however, they require considerably stronger assumptions and present
greater challenges in ensuring valid transportability. To date, MAIC and
STC have been the only PAIC methods available to conduct unanchored
single-arm PAICs. However, as highlighted in preceding sections, they typically
cannot transport estimates beyond the comparator population except in
special cases where \(\Delta_{BC}\) is directly transportable (Section
6.4).

To address these limitations, Chandler and Ishak introduced
multilevel unanchored meta-regression for single-arm indirect comparisons (ML-UMR-SA), an adaptation of ML-NMR
designed for fully unanchored PAICs.\textsuperscript{84,85} ML-UMR-SA enables
synthesis of two or more treatments and allows relative effects to be
estimated in pre-specified target populations, subject to additional
identifying assumptions. Central among these is the shared prognostic factor
assumption (SPFA), which requires all covariates to have the same
effect on outcomes across treatments---implying an absence of
conditional effect modification under the structural assumptions of the
model. SPFA is conceptually analogous to SEMA, which allows effect
modification to cancel between active arms in conditional anchored
contrasts under scale alignment; however, SPFA is generally more
restrictive and more difficult to justify
empirically.\textsuperscript{84,85}

In principle, SPFA may be relaxed (i.e., allow for effect modification
between active arms) by incorporating subgroup-specific comparator
information, multiple comparator studies, or external clinical
knowledge.\textsuperscript{83-85} In practice, however, obtaining the
data required to relax this assumption is challenging due to limited
reporting of comparator information. While ML-UMR-SA broadens the scope of
transportability in unanchored settings, it remains subject to strong
limitations inherent to unanchored comparisons. Further methodological
work is needed to characterize when these assumptions are plausible and
how they can be assessed in HTA applications.

\hypertarget{implications-for-paics-and-hta}{%
\section{\texorpdfstring{ Implications for PAICs and
HTA}{ Implications for PAICs and HTA}}\label{implications-for-paics-and-hta}}

The results of this work have several implications for the
interpretation and application of relative treatment effects in HTA,
across a range of evidence synthesis methods, including PAICs. In
commonly used MAIC and STC analyses targeting marginal estimands on
non-collapsible effect scales, the estimated active-to-active effect is
generally defined in the comparator trial population and depends on that
population's baseline risk and covariate distribution. As a result, such
estimates are not directly transportable back to the index population,
nor to other target populations, without invoking additional assumptions
beyond those used to justify the initial adjustment.

This population dependence helps explain why MAICs or STCs conducted by
different sponsors may lead to differing results or
conclusions.\textsuperscript{19,86} In practice, each sponsor typically
uses individual-level data from its own trial and adjusts to a different
comparator study population, so the resulting active-to-active estimates
are defined in different populations for each sponsor. Apparent
discrepancies across submissions therefore need not reflect analytic
error, but rather differences in the underlying estimand, underscoring
the importance of clearly identifying the population in which a PAIC
estimate is defined.

The SEMA remains an important component of PAIC methodology,
particularly for approaches such as ML-NMR that aim to estimate
treatment effects in pre-specified target populations. However, our
findings indicate that its role is more limited than is often stated in
current HTA guidance, which frequently treats SEMA as sufficient to
justify direct transportability of active-to-active effects (e.g., NICE
DSU TSD 18; EU HTA Joint Clinical Assessment guidance). In pairwise MAIC
and STC, estimation of active-to-active effects in the comparator
population does not generally require SEMA, because these methods rely
on conditional transport of index-trial effects and typically target
marginal estimands rather than conditional effects. SEMA becomes
relevant for MAIC and STC only in special cases where direct
transportability of the marginal estimand is assumed, such as for
collapsible effect measures with appropriate scale alignment, or if STC
targets a conditional effect (rarely possible in practice). More
broadly, while SEMA may be necessary for certain forms of
transportability, it is not sufficient to guarantee direct
transportability of marginal effects, which also depends on the target
estimand, the collapsibility of the effect measure, and alignment
between the scale of effect modification and the effect scale;
consequently, even when SEMA is plausible, marginal effects may vary
across populations.

The magnitude of population dependence in practice will vary.
Differences in baseline risk, covariate distributions, and the strength
of effect modification and prognostic effects all contribute to
variation in \(\Delta_{BC}\) across populations. In settings with strong
covariate overlap and weak effect modification, transported effects may
be approximately stable, such that
\(\Delta_{BC(AC)} \approx \Delta_{BC(AB)}\). However, this approximation
is context-specific and should not be assumed without empirical support
or sensitivity analysis.

When a connected treatment network is available and modeling assumptions
are plausible, ML-NMR provides a framework for estimating relative
effects in pre-specified target populations and can reduce reliance on
the implicit two-step transport assumption inherent to pairwise MAIC and
STC. The potential for broader transportability properties with ML-NMR,
however, relies on additional assumptions: notably, shared prognostic
effects and SEMA holding across the entire network. In unanchored
settings, transportability challenges are inherently more complex, but
extensions such as ML-UMR-SA represent an important step toward addressing
these challenges by enabling synthesis across multiple studies and
clarifying the assumptions required to estimate effects in
decision-relevant target populations.

Finally, these findings extend beyond PAICs. Relative treatment effects
are frequently transported in HTA more broadly, including when marginal
effect estimates from RCTs, Bucher comparisons, or NMAs are applied
within economic models defined in different populations. For
non-collapsible measures such as hazard ratios and odds ratios, such
practice implicitly assumes stability across baseline risk and joint
covariate distributions---assumptions that may not hold in many
applications, particularly when transporting across populations without
explicit re-standardization. Because this practice is common,
cost-effectiveness estimates from economic models in HTA may often
reflect bias induced by transport steps. Greater clarity regarding
estimands, target populations, and transportability conditions can
therefore improve the transparency and credibility of evidence synthesis
used in HTA decision-making.

\hypertarget{concluding-remarks}{%
\section{\texorpdfstring{ Concluding
Remarks}{ Concluding Remarks}}\label{concluding-remarks}}

This paper examines how relative treatment effects from PAICs are
applied outside the populations in which they are estimated. In
practice, pairwise PAICs often identify an active-to-active effect
defined in the comparator trial population, yet this estimate is
subsequently applied to an index trial or decision-model population with
different baseline risk and covariate distributions. Using illustrative
numerical examples, we show that commonly used PAIC approaches,
including MAIC and STC, typically identify population-specific marginal
effects when targeting non-collapsible measures. While the shared effect
modifier assumption remains central to PAIC methodology, it is not
sufficient on its own to ensure direct transportability across
populations. Viewing PAICs as transport problems rather than purely
adjustment problems clarifies the additional assumptions implicitly
invoked when PAIC-derived effects are applied beyond the comparator
population, and highlights the critical roles of estimand choice, effect
scale, and modeling assumptions.

Our analysis intentionally conditions on successful completion of the
first transport step in pairwise PAICs (i.e., conditional transport of
\(\Delta_{BA}\) to the comparator population)---a step known to be
challenging for marginal estimands\textsuperscript{18,35}---to isolate
the transportability properties of the resulting active-to-active
estimand. Even under this optimistic assumption, we show that commonly
used PAIC estimands, particularly non-collapsible measures, are often
not directly transportable across populations.

Transportability challenges are exacerbated when access to complete IPD
is restricted, as is typical in HTA settings. Limited data availability
constrains assessment of effect modification and population overlap,
increasing reliance on strong and often unverifiable assumptions. Where
feasible, comparative analyses based on complete IPD (e.g., propensity
score--based matching or weighting approaches) can facilitate
transportability by enabling direct population alignment and reducing
dependence on indirect transport assumptions.\textsuperscript{87}
Broader access to IPD across sponsors would therefore strengthen the
basis of comparative effectiveness assessments and mitigate key
limitations of PAICs.

More broadly, transportability is not an intrinsic property of a method
or effect measure, but a consequence of the interaction between estimand
choice, modeling assumptions, and population composition. Making these
elements explicit---by clearly specifying the target estimand, the
population in which it is defined, and the assumptions under which it is
transported---can improve the transparency, interpretability, and
credibility of synthesized evidence used in HTA and related
decision-making contexts.

\hypertarget{references}{%
\section{\texorpdfstring{ References}{ References}}\label{references}}

\hspace*{4mm} 1. Wale JL, Thomas S, Hamerlijnck D, Hollander R. Patients and public
are important stakeholders in health technology assessment but the level
of involvement is low - a call to action. \emph{Res Involv Engagem}. Jan
5 2021;7(1):1. doi:10.1186/s40900-020-00248-9

2. Hogervorst MA, Ponten J, Vreman RA, Mantel-Teeuwisse AK, Goettsch WG.
Real World Data in Health Technology Assessment of Complex Health
Technologies. \emph{Front Pharmacol}. 2022;13:837302.
doi:10.3389/fphar.2022.837302

3. Jaksa A, Arena PJ, Hanisch M, Marsico M. Use of Real-World Evidence
in Health Technology Reassessments Across 6 Health Technology Assessment
Agencies. \emph{Value Health}. Jun 2025;28(6):898-906.
doi:10.1016/j.jval.2025.02.012

4. Hariton E, Locascio JJ. Randomised controlled trials - the gold
standard for effectiveness research: Study design: randomised controlled
trials. \emph{BJOG}. Dec 2018;125(13):1716. doi:10.1111/1471-0528.15199

5. Rothwell PM. Factors that can affect the external validity of
randomised controlled trials. \emph{PLoS Clin Trials}. May 2006;1(1):e9.
doi:10.1371/journal.pctr.0010009

6. Steckler A, McLeroy KR. The importance of external validity. \emph{Am
J Public Health}. Jan 2008;98(1):9-10. doi:10.2105/AJPH.2007.126847

7. Member State Coordination Group on Health Technology Assessment (HTA
CG). Guidance on Validity of Clinical Studies. Accessed 2025,
\url{https://health.ec.europa.eu/document/download/9f9dbfe4-078b-4959-9a07-df9167258772_en?filename=hta_clinical-studies-validity_guidance_en.pdf}

8. Van Spall HG, Toren A, Kiss A, Fowler RA. Eligibility criteria of
randomized controlled trials published in high-impact general medical
journals: a systematic sampling review. \emph{JAMA}. Mar 21
2007;297(11):1233-40. doi:10.1001/jama.297.11.1233

9. Hernán MA, Robins JM. \emph{Causal Inference: What If}. Chapman \&
Hall/CRC; 2020.

10. Stuart EA, Cole SR, Bradshaw CP, Leaf PJ. The use of propensity
scores to assess the generalizability of results from randomized trials.
\emph{Journal of the Royal Statistical Society: Series A}.
2011;174(2):369-386. doi:10.1111/j.1467-985X.2010.00673.x

11. Turner AJ, Sammon C, Latimer N. Transporting Comparative
Effectiveness Evidence Between Countries: Considerations for Health
Technology Assessments. \emph{PharmacoEconomics}. 2024;42(2):165-176.
doi:10.1007/s40273-023-01323-1

12. Murad MH, Katabi A, Benkhadra R, Montori VM. External validity,
generalisability, applicability and directness: a brief primer.
\emph{BMJ Evid Based Med}. Feb 2018;23(1):17-19.
doi:10.1136/ebmed-2017-110800

13. Vuong Q, Metcalfe RK, Ling A, Ackerman B, Inoue K, Park JJ.
Systematic review of applied transportability and generalizability
analyses: A landscape analysis. \emph{Ann Epidemiol}. Apr
2025;104:61-70. doi:10.1016/j.annepidem.2025.03.001

14. Bareinboim E, Pearl J. Causal inference and the data-fusion problem.
\emph{Proceedings of the National Academy of Sciences}.
2016;113(27):7345-7352. doi:10.1073/pnas.1510507113

15. Levy NS, Arena PJ, Jemielita T, et al. Use of transportability
methods for real-world evidence generation: a review of current
applications. \emph{J Comp Eff Res}. Nov 2024;13(11):e240064.
doi:10.57264/cer-2024-0064

16. Dahabreh IJ, Robertson SE, Steingrimsson JA, Stuart EA, Hernan MA.
Extending inferences from a randomized trial to a new target population.
\emph{Stat Med}. Jun 30 2020;39(14):1999-2014. doi:10.1002/sim.8426

17. Dahabreh IJ, Robertson SE, Steingrimsson JA. Learning about treatment effects in a new target population under transportability assumptions for relative effect measures. \emph{Eur J Epidemiol.} 2024 Sep;39(9):957-965. doi: 10.1007/s10654-023-01067-4

18. Remiro-Azócar A, Phillippo, D. M., Welton, N. J., Dias, S., Ades, A.
E., Heath, A., \& Baio, G. Marginal and conditional summary measures:
Transportability and compatibility across studies. \emph{Wiley StatsRef:
Statistics Reference Online}. 2025;

19. Ishak KJ, Chandler C, Liu FF, Klijn S. A Framework for Reliable,
Transparent, and Reproducible Population-Adjusted Indirect Comparisons.
\emph{Pharmacoeconomics}. May 5 2025;doi:10.1007/s40273-025-01503-1

20. Macabeo B, Rotrou T, Millier A, Francois C, Laramee P. The
Acceptance of Indirect Treatment Comparison Methods in Oncology by
Health Technology Assessment Agencies in England, France, Germany,
Italy, and Spain. \emph{Pharmacoecon Open}. Jan 2024;8(1):5-18.
doi:10.1007/s41669-023-00455-6

21. Bucher HC, Guyatt GH, Griffith LE, Walter SD. The results of direct
and indirect treatment comparisons in meta-analysis of randomized
controlled trials. \emph{Journal of Clinical Epidemiology}.
1997;50(6):683-691. doi:10.1016/S0895-4356(97)00049-8

22. Dias S, Welton NJ, Sutton AJ, AE A. NICE DSU Technical Support
Document 2: A generalised linear modelling framework for pair-wise and
network meta-analysis of randomised controlled trials. 2011;

23. Dias S, Sutton AJ, Ades AE, Welton NJ. Evidence synthesis for
decision making 2: a generalized linear modeling framework for pairwise
and network meta-analysis of randomized controlled trials. \emph{Med
Decis Making}. Jul 2013;33(5):607-17. doi:10.1177/0272989X12458724

24. Lumley T. Network meta-analysis for indirect treatment comparisons.
\emph{Stat Med}. Aug 30 2002;21(16):2313-24. doi:10.1002/sim.1201

25. Hoaglin DC, Hawkins N, Jansen JP, et al. Conducting
indirect-treatment-comparison and network-meta-analysis studies: report
of the ISPOR Task Force on Indirect Treatment Comparisons Good Research
Practices: part 2. \emph{Value Health}. Jun 2011;14(4):429-37.
doi:10.1016/j.jval.2011.01.011

26. Ades AE, Welton NJ, Dias S, Phillippo DM, Caldwell DM. Twenty years
of network meta-analysis: Continuing controversies and recent
developments. \emph{Res Synth Methods}. Jan 18
2024;doi:10.1002/jrsm.1700

27. Ishak KJ, Proskorovsky I, Benedict A. Simulation and matching-based
approaches for indirect comparison of treatments.
\emph{Pharmacoeconomics}. Jun 2015;33(6):537-49.
doi:10.1007/s40273-015-0271-1

28. Phillippo DM, Ades AE, Dias S, Palmer S, Abrams KR, Welton NJ.
Methods for population-adjusted indirect comparisons in health
technology appraisal. \emph{Medical Decision Making}.
2018;38(2):200-211. doi:10.1177/0272989X17725740

29. Phillippo DM, Dias S, Elsada A, Ades AE, Welton NJ. Population
Adjustment Methods for Indirect Comparisons: A Review of National
Institute for Health and Care Excellence Technology Appraisals.
\emph{Int J Technol Assess Health Care}. Jan 2019;35(3):221-228.
doi:10.1017/S0266462319000333

30. Caro JJ, Ishak KJ. No head-to-head trial? simulate the missing arms.
\emph{Pharmacoeconomics}. 2010;28(10):957-67.
doi:10.2165/11537420-000000000-00000

31. Signorovitch JE, Sikirica V, Erder MH, et al. Matching-adjusted
indirect comparisons: a new tool for timely comparative effectiveness
research. \emph{Value Health}. Sep-Oct 2012;15(6):940-7.
doi:10.1016/j.jval.2012.05.004

32. Signorovitch JE, Wu EQ, Yu AP, et al. Comparative effectiveness
without head-to-head trials: a method for matching-adjusted indirect
comparisons applied to psoriasis treatment with adalimumab or
etanercept. \emph{Pharmacoeconomics}. 2010;28(10):935-45.
doi:10.2165/11538370-000000000-00000

33. Macabeo B, Quenechdu A, Aballea S, Francois C, Boyer L, Laramee P.
Methods for Indirect Treatment Comparison: Results from a Systematic
Literature Review. \emph{J Mark Access Health Policy}. Jun
2024;12(2):58-80. doi:10.3390/jmahp12020006

34. Phillippo DM, Dias S, Ades AE, Welton NJ. Assessing the performance
of population adjustment methods for anchored indirect comparisons: A
simulation study. \emph{Stat Med}. Dec 30 2020;39(30):4885-4911.
doi:10.1002/sim.8759

35. Remiro-Azocar A. Transportability of model-based estimands in
evidence synthesis. \emph{Stat Med}. Sep 30 2024;43(22):4217-4249.
doi:10.1002/sim.10111

36. International Council for Harmonisation of Technical Requirements
for Pharmaceuticals for Human Use (ICH). Addendum on estimands and
sensitivity analysis in clinical trials to the guideline on statistical
principles for clinical trials E9(R1). 2019;

37. Remiro-Azocar A. Some considerations on target estimands for health
technology assessment. \emph{Stat Med}. Dec 10 2022;41(28):5592-5596.
doi:10.1002/sim.9566

38. Remiro-Azocar A. Target estimands for population-adjusted indirect
comparisons. \emph{Stat Med}. Dec 10 2022;41(28):5558-5569.
doi:10.1002/sim.9413

39. Pearl J, Bareinboim E. External Validity and Transportability: A
Formal Approach. \emph{International Statistical Institute Proceedings
of the 58th World Statistical Congress (Paper 450268)}. 2011;

40. Pearl J, Bareinboim E. External Validity: From Do-Calculus to
Transportability Across Populations. \emph{Statistical Science}.
2014;29(4):579-595. doi:10.1214/14-STS486

41. Rubin DB. Estimating causal effects of treatments in randomized and
nonrandomized studies. \emph{Journal of Educational Psychology}.
1974;66:688--701.

42. Holland PW. Statistics and causal inference. \emph{Journal of the
American Statistical Association}. 1986;81:945--70.

43. van Amsterdam WAC, Elias S, Ranganath R. Causal Inference in
Oncology: Why, What, How and When. \emph{Clin Oncol (R Coll Radiol)}.
Feb 2025;38:103616. doi:10.1016/j.clon.2024.07.002

44. VanderWeele TJ. Commentary: On Causes, Causal Inference, and
Potential Outcomes. \emph{Int J Epidemiol}. Dec 1 2016;45(6):1809-1816.
doi:10.1093/ije/dyw230

45. Phillippo DM R-AA, Heath A, Baio G, Dias S, Ades AE, Welton NJ.
Effect modification and non-collapsibility together may lead to
conflicting treatment decisions: A review of marginal and conditional
estimands and recommendations for decision-making. \emph{Research
Synthesis Methods}. 2025;16(2):323--49.

46. Dahabreh IJ, Haneuse SJA, Robins JM, et al. Study Designs for
Extending Causal Inferences From a Randomized Trial to a Target
Population. \emph{Am J Epidemiol}. Aug 1 2021;190(8):1632-1642.
doi:10.1093/aje/kwaa270

47. Snowden JM, Rose S, Mortimer KM. Implementation of G-computation on
a simulated data set: demonstration of a causal inference technique.
\emph{Am J Epidemiol}. Apr 1 2011;173(7):731-8. doi:10.1093/aje/kwq472

48. Remiro-Azocar A, Heath A, Baio G. Parametric G-computation for
compatible indirect treatment comparisons with limited individual
patient data. \emph{Res Synth Methods}. Nov 2022;13(6):716-744.
doi:10.1002/jrsm.1565

49. Remiro-Azocar A, Heath A, Baio G. Model-based standardization using
multiple imputation. \emph{BMC Med Res Methodol}. Feb 10 2024;24(1):32.
doi:10.1186/s12874-024-02157-x

50. Westreich D, Edwards JK, Lesko CR, Stuart E, Cole SR.
Transportability of Trial Results Using Inverse Odds of Sampling
Weights. \emph{Am J Epidemiol}. Oct 15 2017;186(8):1010-1014.
doi:10.1093/aje/kwx164

51. Ling AY, Jreich R, Montez-Rath ME, et al. Transporting observational
study results to a target population of interest using inverse odds of
participation weighting. \emph{PLoS One}. 2022;17(12):e0278842.
doi:10.1371/journal.pone.0278842

52. Stuart EA, Bradshaw CP, Leaf PJ. Assessing the generalizability of
randomized trial results to target populations. \emph{Prev Sci}. Apr
2015;16(3):475-85. doi:10.1007/s11121-014-0513-z

53. Funk MJ, Westreich D, Wiesen C, Sturmer T, Brookhart MA, Davidian M.
Doubly robust estimation of causal effects. \emph{Am J Epidemiol}. Apr 1
2011;173(7):761-7. doi:10.1093/aje/kwq439

54. Lee D, Yang S, Wang X. Doubly robust estimators for generalizing
treatment effects on survival outcomes from randomized controlled trials
to a target population. \emph{J Causal Inference}. 2022;10(1):415-440.
doi:10.1515/jci-2022-0004

55. Li X, Shen C. Doubly Robust Estimation of Causal Effect: Upping the
Odds of Getting the Right Answers. \emph{Circ Cardiovasc Qual Outcomes}.
Jan 2020;13(1):e006065. doi:10.1161/CIRCOUTCOMES.119.006065

56. Campbell H, Remiro-Azócar A. Doubly robust augmented weighting
estimators for the analysis of externally controlled single-arm trials
and unanchored indirect treatment comparison. \emph{arXiv:250500113v2}.
2025;

57. Phillippo DM, Ades AE, Dias S, Palmer S, Abrams KR, NJ W. NICE DSU
Technical Support Document 18: Methods for population-adjusted indirect
comparisons in submissions to NICE. 2016;

58. Guyot P, Welton NJ, Ouwens MJ, Ades AE. Survival time outcomes in
randomized, controlled trials and meta-analyses: the parallel universes
of efficacy and cost-effectiveness. \emph{Value Health}. Jul-Aug
2011;14(5):640-6. doi:10.1016/j.jval.2011.01.008

59. Freeman SC, Cooper NJ, Sutton AJ, Crowther MJ, Carpenter JR, Hawkins
N. Challenges of modelling approaches for network meta-analysis of
time-to-event outcomes in the presence of non-proportional hazards to
aid decision making: Application to a melanoma network. \emph{Stat
Methods Med Res}. May 2022;31(5):839-861. doi:10.1177/09622802211070253

60. Dias S, Welton NJ, Sutton AJ, Ades AE. Evidence synthesis for
decision making 5: the baseline natural history model. \emph{Med Decis
Making}. Jul 2013;33(5):657-70. doi:10.1177/0272989X13485155

61. Member State Coordination Group on Health Technology Assessment (HTA
CG). Practical Guideline for Quantitative Evidence Synthesis: Direct and
Indirect Comparisons. Accessed 2025,
\url{https://health.ec.europa.eu/document/download/1f6b8a70-5ce0-404e-9066-120dc9a8df75_en?filename=hta_practical-guideline_direct-and-indirect-comparisons_en.pdf}

62. Daniel R, Zhang J, Farewell D. Making apples from oranges: Comparing
noncollapsible effect estimators and their standard errors after
adjustment for different covariate sets. \emph{Biom J}. Mar
2021;63(3):528-557. doi:10.1002/bimj.201900297

63. Bénédicte Colnet JJ, Gaël Varoquaux, Erwan Scornet. Risk ratio, odds
ratio, risk difference... Which causal measure is easier to generalize?
\emph{arXiv:230316008}. 2024;

64. Campbell H, Jansen J. Hidden in Plain Sight: How Non-Collapsibility
Biases Treatment Effects in (Network) Meta-Analysis.
\emph{arXiv:260300749}. 2026;

65. Remiro-Azocar A, Heath A, Baio G. Methods for population adjustment
with limited access to individual patient data: A review and simulation
study. \emph{Res Synth Methods}. Nov 2021;12(6):750-775.
doi:10.1002/jrsm.1511

66. Chandler CO, Proskorovsky I. Uncertain about uncertainty in
matching-adjusted indirect comparisons? A simulation study to compare
methods for variance estimation. \emph{Res Synth Methods}. Nov
2024;15(6):1094-1110. doi:10.1002/jrsm.1759

67. Hatswell AJ, Freemantle N, Baio G. The Effects of Model
Misspecification in Unanchored Matching-Adjusted Indirect Comparison:
Results of a Simulation Study. \emph{Value Health}. Jun
2020;23(6):751-759. doi:10.1016/j.jval.2020.02.008

68. Liu Y, He X, Liu J, Wu J. Is the Use of Unanchored Matching-Adjusted
Indirect Comparison Always Superior to Naive Indirect Comparison on
Survival Outcomes? A Simulation Study. \emph{Appl Health Econ Health
Policy}. Jul 2025;23(4):693-704. doi:10.1007/s40258-025-00952-1

69. Ren S, Ren S, Welton NJ, Strong M. Advancing unanchored simulated
treatment comparisons: A novel implementation and simulation study.
\emph{Res Synth Methods}. Jul 2024;15(4):657-670. doi:10.1002/jrsm.1718

70. Zhang L, Bujkiewicz S, Jackson D. Four alternative methodologies for
simulated treatment comparison: How could the use of simulation be
re-invigorated? \emph{Res Synth Methods}. Mar 2024;15(2):227-241.
doi:10.1002/jrsm.1681

71. Jiang Z, Jialing Liu, Weili He, Joseph Cappelleri, Satrajit
Roychoudhury, Yong Chen, and Haitao Chu. The Hazards of Using Hazard
Ratios from Proportional Hazard Models in Indirect Treatment
Comparisons. \emph{Research Synthesis Methods}. 2026;17(3):483-497.

72. Cronin A, Tian L, Uno H. strmst2 and strmst2pw: New commands to
compare survival curves using the restricted mean survival time.
\emph{The Stata Journal}. 2016;16(3):702--716.

73. Zhong Y, Zhao O, Zhang B, Yao B. Adjusting for covariates in
analysis based on restricted mean survival times. \emph{Pharm Stat}. Jan
2022;21(1):38-54. doi:10.1002/pst.2151

74. Conner SC, Sullivan LM, Benjamin EJ, LaValley MP, Galea S, Trinquart
L. Adjusted restricted mean survival times in observational studies.
\emph{Stat Med}. Sep 10 2019;38(20):3832-3860. doi:10.1002/sim.8206

75. Didelez V, Stensrud MJ. On the logic of collapsibility for causal
effect measures. \emph{Biom J}. Feb 2022;64(2):235-242.
doi:10.1002/bimj.202000305

76. Lu WE, Ni A. Causal effect estimation on restricted mean survival
time under case-cohort design via propensity score stratification.
\emph{Lifetime Data Anal}. Oct 2025;31(4):898-931.
doi:10.1007/s10985-025-09667-w

77. Ni A, Lin Z, Lu B. Stratified Restricted Mean Survival Time Model
for Marginal Causal Effect in Observational Survival Data. \emph{Ann
Epidemiol}. Dec 2021;64:149-154. doi:10.1016/j.annepidem.2021.09.016

78. Hernan MA. The hazards of hazard ratios. \emph{Epidemiology}. Jan
2010;21(1):13-5. doi:10.1097/EDE.0b013e3181c1ea43

79. Huitfeldt A, Stensrud MJ, Suzuki E. On the collapsibility of
measures of effect in the counterfactual causal framework. \emph{Emerg
Themes Epidemiol}. 2019;16:1. doi:10.1186/s12982-018-0083-9

80. Sjolander A, Dahlqwist E, Zetterqvist J. A Note on the
Noncollapsibility of Rate Differences and Rate Ratios.
\emph{Epidemiology}. May 2016;27(3):356-9.
doi:10.1097/EDE.0000000000000433

81. Phillippo DM, Dias S, Ades AE, Welton NJ. Multilevel network
meta-regression for population-adjusted treatment comparisons.
\emph{Journal of the Royal Statistical Society: Series A}.
2020;183(3):1189-1210. doi:10.1111/rssa.12579

82. Chandler C, Ishak J. Surviving unanchored indirect comparisons: an
extension of multilevel unanchored meta-regression (ML-UMR) for survival
analyses. Poster presented at: ISPOR---The Professional Society for
Health Economics and Outcomes Research; 2026 May 17--20; Philadelphia,
PA, USA. Poster MSR131. 2026;

83. Campbell H, Margossian C, Jansen J, Gustafson P. Don't Disregard the
Data for Lack of a Likelihood: Bayesian Synthetic Likelihood for
Enhanced Multilevel Network Meta-Regression. \emph{arXiv:260311019}.
2026;

84. Chandler C, Ishak J. Anchors Away: Navigating Unanchored Indirect
Comparisons With Multilevel Unanchored Meta-Regression. \emph{Value in
Health}. 2025;28(12)(1):S498.

85. Chandler C, Ishak J. Anchors Away: Navigating Single-Arm Indirect Comparisons with Multilevel Unanchored Meta-Regression (ML-UMR-SA). \emph{arXiv}. 2026;doi:10.48550/arXiv.2606.20341

86. Jiang Z, Liu, J., Alemayehu, D., Cappelleri, J. C., Abrahami, D.,
Chen, Y., Chu, H. A critical assessment of matching-adjusted indirect
comparisons in relation to target populations. \emph{Research Synthesis
Methods}. 2025;16(3):569--574.

87. Faria R, Hernandez Alava M, Manca A, AJ W. NICE DSU Technical
Support Document 17: The use of observational data to inform estimates
of treatment effectiveness in technology appraisal: methods for
comparative individual patient data. 2015;

\begin{Backmatter}

\paragraph{Acknowledgments}
The authors thank Caroline Cole, Ruth Sharf, and Richard Leason for editorial and graphical design support on this manuscript. The authors also used OpenAI's ChatGPT 5.5 to assist with proofreading and copyediting of manuscript text. The authors are also grateful to the anonymous peer reviewers for their incisive feedback, which helped refine the manuscript’s conceptual framing and strengthen its discussion of newer population-adjustment methods, including ML-NMR.

\paragraph{Funding Statement}
This study and article were funded by Thermo Fisher Scientific. The authors are employees of Thermo Fisher Scientific.

\paragraph{Competing Interests}
The authors declare none.

\paragraph{Data Availability Statement}
Simulated examples are presented in this article. The associated R code used to generate and analyze the data is available in the supplement.

\paragraph{Ethical Standards}
Not applicable. This study did not involve human participants, animal subjects, or identifiable personal data.

\paragraph{Author Contributions}
Conor O. Chandler: Conceptualization, methodology, writing – original draft, writing – reviewing and editing, visualization, formal analysis; K. Jack Ishak: Conceptualization, methodology, writing – reviewing and editing.

\end{Backmatter}

\newpage

\noindent \textbf{Supplementary Material for ``Reframing Population-Adjusted
Indirect Comparisons as a Transportability Problem: An Estimand-Based
Perspective and Implications for Health Technology Assessment''}

\hypertarget{appendix-a-proposition-and-proofs-of-direct-transportability-conditions}{%
\subsection{Appendix A --- Proposition and Proofs of Direct
Transportability
Conditions}\label{appendix-a-proposition-and-proofs-of-direct-transportability-conditions}}

\hypertarget{conditions-for-direct-transportability-of-conditional-effects}{%
\subsubsection{Conditions for Direct Transportability of Conditional
Effects}\label{conditions-for-direct-transportability-of-conditional-effects}}

\vspace{2mm}
\textbf{Proposition A1 (Direct transportability of conditional contrasts
on the linear predictor scale).}\\
Let \(P\) denote a population with covariate distribution \(f_{P}(x)\)
with baseline covariates \(X\). Consider three treatments \(A\) (common
comparator), \(B\), and \(C\). Suppose the conditional mean of the
potential outcome under each treatment \(t \in \{ A,B,C\}\) has the
following form

\begin{equation}
g\!\left(
\mathbb{E}\!\left[Y^{t}\mid X=x\right]
\right)
=
m(x)
+
\phi(x)
+
\delta_t
\tag{A1}
\end{equation}

where

\begin{itemize}
\item
  \(g( \cdot )\) is a model link function defining the linear predictor
  scale,
\item
  $m(x) = \beta_{0} + Z(X)^{\top}\beta^{\mathrm{PF}}$ is a baseline prognostic
  function, representing how covariates $x$ influence the outcome
  independent of treatment (i.e., common to all treatments),
\item
  \(\delta_{t}\) is a treatment-specific main effect on the \(g\)-scale
  (with \(\delta_{A} \equiv 0\)), and
\item
  \(\phi(x) = \left( {Z(X)}^{T}\beta^{\mathrm{EM}} \right)\mathbf{1}(t \neq A)\)
  is an effect modification function on the $g$-scale, describing how
  covariates $x$ modify the effect of each active treatment relative to
  the common comparator \(A\)
\end{itemize}

Assume the SEMA holds on the \(g\)-scale, i.e. the same \(\phi(x)\)
applies to both \(B\) and \(C\) (relative to \(A\)). No parametric or
linearity assumptions are imposed on \(m(x)\) or \(\phi(x)\). For
instance \(m(x)\) may capture any type of baseline risk differences
across patients (e.g., interactions, higher-order terms). The only
structural assumptions required are additivity on the g-scale and
sharing of \(\phi(x)\) across active treatments under SEMA (i.e.,
\(\beta_{B}^{\mathrm{EM}} = \beta_{C}^{\mathrm{EM}}\)).

Let \(h( \cdot )\) be a transformation defining the conditional contrast
of interest on the outcome (estimand) scale. Define the individual conditional
active--active contrast at covariate value \(x\) by

\begin{equation}
\Delta_{BC}^{\mathrm{Cond}}(x)
:=
h\!\left(g^{-1}\!\bigl(\eta_{B}(x)\bigr)\right)
-
h\!\left(g^{-1}\!\bigl(\eta_{C}(x)\bigr)\right)
\tag{A2}
\end{equation}

where

\begin{equation}
\begin{aligned}
\eta_{B}(x) &:= m(x) + \phi(x) + \delta_{B}, \\
\eta_{C}(x) &:= m(x) + \phi(x) + \delta_{C}.
\end{aligned}
\tag{A3}
\end{equation}

\noindent \textbf{(i) Direct transportability of conditional contrasts under scale
alignment.}\\
If \(h \circ g^{- 1}\) is the identity function on the range of the
linear predictor (i.e. \(h(g^{- 1}(u)) = u\) for all \(u\) in the range
of \(\eta_{t}(X)\)), then for every \(x\)

\[
\Delta_{BC}^{\mathrm{Cond}}(x)
=
\delta_{B} - \delta_{C},
\]

which is constant in \(x\). Equivalently, the effect measure operates on
the same scale as the model's linear predictor. Consequently, for any
population \(P\),

\[
\mathbb{E}_{X \sim P}\!\left[\Delta_{BC}^{\mathrm{Cond}}(X)\right]
= \delta_{B} - \delta_{C},
\]

as the \(\delta\)'s do not depend on \(x\). That is, the conditional
active--active contrast is \textbf{directly transportable} (i.e.,
invariant to the population's covariate distribution) under SEMA.

\vspace{2mm}

\noindent \textbf{(ii) Failure of direct transportability without scale
alignment.}\\
If \(h \circ g^{- 1}\) is not the identity function (so that the
estimand does not operate on the same linear-predictor scale), then in
general \(\Delta_{BC}^{\text{Cond}}(x)\) depends on \(x\) via \(m(x)\) and
\(\phi(x)\), and its population average may vary across populations with
different \(f_{P}(x)\). Thus, SEMA alone does not guarantee direct
transportability in that case.

\vspace{3mm}

\noindent \textbf{Proof of Proposition A1.}

\noindent From (A3),

\[
\begin{aligned}
\eta_{B}(x) - \eta_{C}(x)
&= \bigl(m(x)  + \phi(x) + \delta_{B}\bigr)
   - \bigl(m(x) + \phi(x) + \delta_{C}\bigr) \\
&= \delta_{B} - \delta_{C},
\end{aligned}
\]

which does not depend on \(x\).

Under the scale-alignment condition \(h \circ g^{- 1} = I( \cdot )\),
where \(I(u) = u\) denotes the identity function, we have

\[
\begin{aligned}
\Delta_{BC}^{\mathrm{Cond}}(x)
&= h\!\left(g^{-1}\!\bigl(\eta_{B}(x)\bigr)\right)
   - h\!\left(g^{-1}\!\bigl(\eta_{C}(x)\bigr)\right) \\
&= \eta_{B}(x) - \eta_{C}(x) \\
&= \delta_{B} - \delta_{C},
\end{aligned}
\]

so \(\Delta_{BC}^{\text{Cond}}(x)\) is constant in \(x\).

Averaging over any covariate distribution \(f_{P}(x)\) therefore yields
\(\delta_{B} - \delta_{C}\), proving direct transportability.

If \(h \circ g^{- 1}\) is not the identity function, then the difference

\[
h\!\left(g^{-1}\!\bigl(\eta_{B}(x)\bigr)\right)
-
h\!\left(g^{-1}\!\bigl(\eta_{C}(x)\bigr)\right)
\]

cannot in general be written solely as a function of
\(\eta_{B}(x) - \eta_{C}(x)\) , and therefore does not simplify to
\(\delta_{B} - \delta_{C}\); it will typically depend on the individual
values of the linear predictors \(\eta_{B}(x)\) and \(\eta_{C}(x)\), and
therefore on the full joint distribution of covariates through both the
baseline prognostic function \(m(x)\) and the effect-modification
function \(\phi(x)\), rather than collapsing to a function of the simple
difference \(\delta_{B} - \delta_{C}\). Therefore, averaging this
contrast over different populations with different covariate
distributions \(f_{P}(x)\) can produce different values of the
population-averaged contrast, so SEMA by itself is insufficient for
transportability. \(\blacksquare\)

\vspace{3mm}

\noindent \textbf{Remark A1.} Direct transportability of the conditional
active-to-active contrast requires both SEMA and scale alignment. More
generally, relaxing SEMA to allow treatment-specific effect-modification
functions \(\phi_{B}(x)\) and \(\phi_{C}(x)\),

\[
\mathbb{E}_{X \sim P}\!\left[
\Delta_{BC}^{\mathrm{Cond}}(X)
\right]
=
\mathbb{E}_{X \sim P}\!\left[
h\!\left(
g^{-1}\!\left[
m(X)+\phi_{B}(X)+\delta_{B}
\right]
\right)
-
h\!\left(
g^{-1}\!\left[
m(X)+\phi_{C}(X)+\delta_{C}
\right]
\right)
\right].
\]

This expression generally depends on the target population through the
distribution of \(X\). Under scale alignment
\(\left( h \circ g^{- 1} = I \right)\), it simplifies to

\[
\mathbb{E}_{X \sim P}\!\left[
\Delta_{BC}^{\mathrm{Cond}}(X)
\right]
=
(\delta_{B}-\delta_{C})
+
\mathbb{E}_{X \sim P}\!\left[
\phi_{B}(X)-\phi_{C}(X)
\right].
\]

Thus, differential effect modification (i.e., SEMA violation) and scale
misalignment are distinct sources of failure of direct transportability
in indirect treatment comparisons.

\vspace{2mm}

\noindent \textbf{Table A1.} Examples for A1

{\footnotesize
\begin{longtable}{@{}L{0.18\linewidth} L{0.16\linewidth} L{0.17\linewidth} L{0.12\linewidth} L{0.25\linewidth}@{}}
\toprule
\textbf{Conditional effect measure} &
\textbf{Outcome model link $g(\cdot)$} &
\textbf{Estimand transformation $h(\cdot)$} &
$h\circ g^{-1}=I$? &
\textbf{Direct transportability of $\boldsymbol{\Delta}_{BC}^{\mathrm{Cond}}$ under SEMA} \\
\midrule
\endfirsthead

\toprule
\textbf{Conditional effect measure} &
\textbf{Outcome model link $g(\cdot)$} &
\textbf{Estimand transformation $h(\cdot)$} &
$h\circ g^{-1}=I$? &
\textbf{Direct transportability of $\boldsymbol{\Delta}_{BC}^{\mathrm{Cond}}$ under SEMA} \\
\midrule
\endhead

\midrule
\multicolumn{5}{r}{\footnotesize Continued on next page} \\
\midrule
\endfoot

\bottomrule
\endlastfoot

Mean difference & Identity & Identity & Yes & Yes \\
Log odds ratio & Logit & Logit & Yes & Yes \\
Risk difference & Logit & Identity & No & No, due to scale misalignment \\
Restricted mean survival time difference & Log hazard & RMST & No & No, due to scale misalignment \\
\end{longtable}
}

\hypertarget{conditions-for-direct-transportability-of-marginal-effects}{%
\subsubsection{Conditions for Direct Transportability of Marginal
Effects}\label{conditions-for-direct-transportability-of-marginal-effects}}

In addition to scale alignment and SEMA, direct transportability of
marginal effects requires that the measure be directly collapsible ---
that is, the marginal effect must equal the population-average
conditional effect. Proposition A2 formalizes this: when
\(h \circ g^{- 1}\) is the identity function and the marginal measure is
directly collapsible, marginal contrasts coincide with the conditional
contrasts \(\delta_{B} - \delta_{C}\) and are therefore directly
transportable across populations. Conversely, if a measure is
non-collapsible (e.g., marginal odds ratios), then direct
transportability of marginal effects typically fails even when
conditional effects are directly transportable.

\vspace{2mm}
\noindent \textbf{Proposition A2 (Direct transportability of marginal contrasts
under SEMA and direct collapsibility).}\\
Maintain the setup of Proposition A1 and definitions (A1)--(A3). Let
\(\Delta_{BC(P)}^{\mathrm{Marg}}\) denote the marginal contrast of treatments
\(B\)and \(C\) in population \(P\), defined by the estimand

\[
\Delta_{BC(P)}^{\mathrm{Marg}}
=
\Psi\!\left(
\mathbb{E}_{X \sim P}\!\left[Y^{B}\right],
\,
\mathbb{E}_{X \sim P}\!\left[Y^{C}\right]
\right),
\]

for a contrast function \(\Psi( \cdot , \cdot )\) corresponding to the
marginal effect measure of interest (e.g., risk difference).

Suppose both of the following hold:

\begin{enumerate}
\def\labelenumi{\arabic{enumi}.}
\item
  \textbf{Scale alignment}: \(h \circ g^{- 1} = I( \cdot )\), so
  Proposition A1 applies, implying \(\Delta_{BC}^{\mathrm{Cond}}(x) = \delta_{B} - \delta_{C}\) for all \(x\).

\def\labelenumi{\arabic{enumi}.}
\setcounter{enumi}{1}
\item
  \textbf{Direct collapsibility of the marginal measure} \(\Psi\): for
  any population \(P\) and any conditional contrast that is constant in
  \(x\), the marginal contrast equals that constant. Equivalently, when
  conditional contrasts are constant in \(x\) the marginal contrast is
  equal to the population-average conditional contrast.
\end{enumerate}

\[
\Delta_{BC(P)}^{\mathrm{Marg}}
=
\mathbb{E}_{X \sim P}\!\left[\Delta_{BC}^{\mathrm{Cond}}(X)\right]
=
\Delta_{BC(P)}^{\mathrm{Cond}}.
\]

Then for every population \(P\),

\[
\Delta_{BC(P)}^{\mathrm{Marg}}
=
\Delta_{BC(P)}^{\mathrm{Cond}}
=
\delta_{B} - \delta_{C},
\]

so the marginal active--active contrast is directly transportable across
populations under SEMA.

\vspace{2mm}

\noindent \textbf{Proof of Proposition A2.}\\
By (1) and Proposition A1,

\[
\Delta_{BC}^{\mathrm{Cond}}(x)
=
\delta_{B} - \delta_{C},
\quad \forall x.
\]

By the direct-collapsibility assumption (2), when conditional contrasts
are constant in x, the marginal contrast equals the population-average
conditional contrast. Hence

\[
\begin{aligned}
\Delta_{BC(P)}^{\mathrm{Marg}}
&=
\Delta_{BC(P)}^{\mathrm{Cond}} \\
&=
\eta_{B}(x) - \eta_{C}(x) \\
&=
\bigl(m(x)  + \phi(x) + \delta_{B}\bigr)
-
\bigl(m(x)  + \phi(x) + \delta_{C}\bigr) \\
&=
\delta_{B} - \delta_{C}.
\end{aligned}
\]

for any population \(P\), establishing direct transportability of the
marginal contrast. \(\blacksquare\)

\vspace{3mm}

\noindent \textbf{Table A2.} Examples for A2

{\footnotesize
\begin{longtable}{@{}L{0.17\linewidth} 
                    L{0.13\linewidth} 
                    L{0.15\linewidth} 
                    L{0.10\linewidth} 
                    L{0.12\linewidth} 
                    L{0.18\linewidth}@{}}
\toprule
\textbf{Marginal effect measure} &
\textbf{Outcome model link $g(\cdot)$} &
\textbf{Estimand transformation $h(\cdot)$} &
$h\circ g^{-1}=I$? &
\textbf{Directly collapsible?} &
\textbf{Direct transportability of $\Delta_{BC}^{\mathrm{Marg}}$ under SEMA} \\
\midrule
\endfirsthead

\toprule
\textbf{Marginal effect measure} &
\textbf{Outcome model link $g(\cdot)$} &
\textbf{Estimand transformation $h(\cdot)$} &
$h\circ g^{-1}=I$? &
\textbf{Directly collapsible?} &
\textbf{Direct transportability of $\Delta_{BC}^{\mathrm{Marg}}$ under SEMA} \\
\midrule
\endhead

\midrule
\multicolumn{6}{r}{\footnotesize Continued on next page} \\
\midrule
\endfoot

\bottomrule
\endlastfoot

Mean difference & Identity & Identity & Yes & Yes & Yes \\
Odds ratio & Logit & Logit & Yes & No & No \\
Risk difference & Logit & Identity & No & Yes & No \\
RMST difference & Log hazard & RMST & No & Yes & No \\
\end{longtable}
}

\hypertarget{appendix-b-proposition-and-proofs-of-direct-collapsibility-conditions}{%
\subsection{Appendix B --- Proposition and Proofs of Direct
Collapsibility
Conditions}\label{appendix-b-proposition-and-proofs-of-direct-collapsibility-conditions}}

\hypertarget{contrast-of-directly-collapsible-contrasts-is-itself-directly-collapsible}{%
\subsubsection{Contrast of Directly Collapsible Contrasts is Itself
Directly
Collapsible}\label{contrast-of-directly-collapsible-contrasts-is-itself-directly-collapsible}}

In anchored ITCs, the active--active contrast is typically constructed
on an additive scale as a difference of two contrasts (e.g.,
\(\Delta_{BC} = \Delta_{BA} - \Delta_{CA}\)). The following proposition formalizes that if each contrast is directly collapsible, then their difference is also directly collapsible.

\vspace{2mm}

\noindent \textbf{Proposition B0 (Direct collapsibility is preserved under
differences of directly collapsible contrasts).}\\
\noindent Fix a population \(P\) with covariate distribution \(f_{P}(x)\). Let
\(\Delta_{BA}\) and \(\Delta_{CA}\) denote two treatment contrasts
defined on a common additive effect scale (e.g., mean difference).
Suppose each contrast has a conditional effect \(\Delta_{tA}^{\mathrm{Cond}}(x)\)
and a marginal effect \(\Delta_{tA(P)}^{\mathrm{Marg}}\) (for
\(t \in \{ B,C\}\)).

\vspace{2mm}

Assume that both contrasts are directly collapsible in \(P\),
i.e.,

\begin{equation}
\begin{aligned}
\Delta_{BA(P)}^{\mathrm{Marg}}
&= \mathbb{E}_{X \sim P}\!\left[\Delta_{BA}^{\mathrm{Cond}}(X)\right], \\
\Delta_{CA(P)}^{\mathrm{Marg}}
&= \mathbb{E}_{X \sim P}\!\left[\Delta_{CA}^{\mathrm{Cond}}(X)\right].
\end{aligned}
\tag{B0.1}
\end{equation}

Define the anchored active--active contrast on the same additive scale
by

\begin{equation}
\begin{aligned}
\Delta_{BC(P)}^{\mathrm{Marg}}
&:=
\Delta_{BA(P)}^{\mathrm{Marg}}
-
\Delta_{CA(P)}^{\mathrm{Marg}}, \\
\Delta_{BC}^{\mathrm{Cond}}(x)
&:=
\Delta_{BA}^{\mathrm{Cond}}(x)
-
\Delta_{CA}^{\mathrm{Cond}}(x).
\end{aligned}
\tag{B0.2}
\end{equation}

Then \(\Delta_{BC}\) is directly collapsible in \(P\), i.e.,

\begin{equation}
\Delta_{BC(P)}^{\mathrm{Marg}}
=
\mathbb{E}_{X \sim P}\!\left[\Delta_{BC}^{\mathrm{Cond}}(X)\right].
\tag{B0.3}
\end{equation}

\noindent \textbf{Proof of Proposition B0.}\\
By the definition in (B0.2) and the direct-collapsibility assumptions in
(B0.1),

\[
\Delta_{BC(P)}^{\mathrm{Marg}}
=
\Delta_{BA(P)}^{\mathrm{Marg}}
-
\Delta_{CA(P)}^{\mathrm{Marg}}
=
\mathbb{E}\!\left[\Delta_{BA}^{\mathrm{Cond}}(X)\right]
-
\mathbb{E}\!\left[\Delta_{CA}^{\mathrm{Cond}}(X)\right].
\]

Linearity of expectation implies

\[
\mathbb{E}\!\left[\Delta_{BA}^{\mathrm{Cond}}(X)\right]
-
\mathbb{E}\!\left[\Delta_{CA}^{\mathrm{Cond}}(X)\right]
=
\mathbb{E}\!\left[\Delta_{BA}^{\mathrm{Cond}}(X)
-
\Delta_{CA}^{\mathrm{Cond}}(X)\right]
=
\mathbb{E}\!\left[\Delta_{BC}^{\mathrm{Cond}}(X)\right],
\]

where the last equality uses the definition of \(\Delta_{BC}^{\mathrm{Cond}}(x)\)
in (B0.2). This establishes (B0.3). \(\blacksquare\)

\hypertarget{contrast-induced-direct-collapsibility-under-sema-and-scale-alignment}{%
\subsubsection{Contrast-Induced Direct Collapsibility Under SEMA and
Scale
Alignment}\label{contrast-induced-direct-collapsibility-under-sema-and-scale-alignment}}

Unlike standard collapsibility, which is an intrinsic property of an
effect measure (reflecting single treatment contrasts calculated
directly within a given population), the collapsibility results for
anchored active--active contrasts derived here rely on structural
assumptions---most notably SEMA and scale alignment---and therefore
should be interpreted as conditional or induced collapsibility rather
than a general measure-level property. Propositions B1--B3 show how the
properties of direct collapsibility can be induced for anchored
active--active contrasts under SEMA and scale alignment. Throughout,
this induced notion of direct collapsibility refers to equivalence of
the marginal and conditional anchored active--active estimands and does
not imply that the underlying within-population contrasts are
intrinsically directly collapsible.

\vspace{2mm}

\noindent \textbf{Proposition B1 --- Contrast-induced direct collapsibility under
SEMA, scale alignment, and collapsibility}

\noindent In this proposition, we explore the direct collapsibility of
active--active marginal contrasts under SEMA. Maintain the setup and
notation of Propositions A1 and A2. In particular, let \(g( \cdot )\)
denote the link function defining the linear predictor scale, and
suppose the conditional mean of the potential outcome under each
treatment \(t \in \{ A,B,C\}\) satisfies

\begin{equation}
g\!\left(\mathbb{E}\!\left[Y^{t} \mid X = x\right]\right)
=
m(x)  + \phi(x) + \delta_{t}\,\mathbf{1}\{t \neq A\},
\tag{B1}
\end{equation}

with \(\delta_{A} \equiv 0\), where \(m(x)\) is a baseline prognostic
function and \(\phi(x)\) is a shared effect-modification function for
the active treatments \(B\) and \(C\) (i.e., SEMA holds on the
\(g\)-scale).

Let \(h( \cdot )\) define the effect measure of interest, and define the
conditional active--active contrast by

\begin{equation}
\Delta_{BC}^{\mathrm{Cond}}(x)
\:=
h\!\left(g^{-1}\!\bigl(\eta_{B}(x)\bigr)\right)
-
h\!\left(g^{-1}\!\bigl(\eta_{C}(x)\bigr)\right),
\tag{B2}
\end{equation}

where \(\eta_{t}(x) = g(\mathbb{E\lbrack}Y^{t} \mid X = x\rbrack)\).

\vspace{2mm}
\noindent Assume the following:

\begin{enumerate}
\def\labelenumi{\arabic{enumi}.}
\item
  \textbf{Scale alignment.}\\
  \(h \circ g^{- 1} = I( \cdot )\) on the relevant range of the linear
  predictor.
\item
  \textbf{Marginalization of BC estimand via averaging of conditional
  contrasts.}\\
  Following the defining property of collapsibility, the marginal
  active--active contrast \(\Delta_{BC(P)}^{\mathrm{Marg}}\) in any population
  \(P\) can be expressed as a weighted average of the conditional
  contrasts:
\end{enumerate}

\begin{equation}
\Delta_{BC(P)}^{\mathrm{Marg}}
=
\int \Delta_{BC}^{\mathrm{Cond}}(x)\, w_{P}(x)\, dx,
\tag{B3}
\end{equation}

for some weights $w_{P}(x) \ge 0$ satisfying
\[
\int w_{P}(x)\, dx = 1.
\]

\noindent Then the following results hold:

\noindent (i) \textbf{Constant conditional active--active contrast.}\\
Under SEMA and scale alignment (A1),

\begin{equation}
\Delta_{BC}^{\mathrm{Cond}}(x)
=
\delta_{B} - \delta_{C},
\quad \text{for all } x.
\tag{B4}
\end{equation}

\noindent (ii) \textbf{Contrast-induced direct collapsibility and direct
transportability of the marginal contrast.}\\
For every population \(P\),

\begin{equation}
\Delta_{BC(P)}^{\mathrm{Marg}}
=
\delta_{B} - \delta_{C}.
\tag{B5}
\end{equation}

so the marginal active--active contrast exhibits direct collapsibility
(i.e., marginal and conditional effect equivalence) and is invariant to
the population covariate distribution.

\vspace{2mm}

\noindent \textbf{Proof of Proposition B1}

\noindent By scale alignment, \(h\left( g^{- 1}(u) \right) = I(u) = u\), so the
conditional contrast in (B2) reduces to

\[
\Delta_{BC}^{\mathrm{Cond}}(x)
=
\eta_{B}(x) - \eta_{C}(x).
\]

From (B1),

\[
\eta_{B}(x)
=
m(x)  + \phi(x) + \delta_{B},
\qquad
\eta_{C}(x)
=
m(x)  + \phi(x) + \delta_{C},
\]

and therefore

\[
\Delta_{BC}^{\mathrm{Cond}}(x)
=
\bigl(m(x)  + \phi(x) + \delta_{B}\bigr)
-
\bigl(m(x)  + \phi(x)+ \delta_{C}\bigr)
=
\delta_{B} - \delta_{C},
\]

which does not depend on \(x\), establishing (B4).

By the collapsibility assumption (B3),

\[
\begin{aligned}
\Delta_{BC(P)}^{\mathrm{Marg}}
&=
\int \Delta_{BC}^{\mathrm{Cond}}(x)\, w_{P}(x)\, dx \\
&=
(\delta_{B} - \delta_{C})
\int w_{P}(x)\, dx
=
\delta_{B} - \delta_{C},
\end{aligned}
\]

since the weights integrate to one. This holds for any population \(P\),
proving (B5). \(\blacksquare\)

\hypertarget{contrast-induced-direct-collapsibility-risk-ratio-as-an-example}{%
\subsubsection{Contrast-Induced Direct Collapsibility: Risk Ratio as an
Example}\label{contrast-induced-direct-collapsibility-risk-ratio-as-an-example}}

Risk ratios provide a useful illustration of the distinction between
collapsibility and direct collapsibility. Marginal risk ratios can be
expressed as weighted averages of conditional risk ratios and are
therefore collapsible in a general sense. However, the marginal log risk
ratio does not generally equal the population-average conditional log
risk ratio when effect modification is present, so direct collapsibility
does not hold for anchored contrasts (e.g., \hspace{0pt}\(\Delta_{BA}\)
and \hspace{0pt}\(\Delta_{CA}\)). Under SEMA on the log-risk scale,
effect-modification terms cancel in active--active indirect comparisons,
yielding a constant conditional contrast for B vs. C; in this special
case, the marginal and population-average conditional log risk ratios
coincide, yielding assumption-induced direct collapsibility for
\hspace{0pt}\(\Delta_{BC}\). A formal proof of this result follows
Proposition B2 and B3.

\vspace{2mm}

\noindent \textbf{Proposition B2 (Non--direct-collapsibility of $\log(\mathrm{RR}_{AB})$ in general).}

\noindent There exist \(\left\{ \mu_{A}(x),\mu_{B}(x) \right\}\) and a covariate
distribution \(f_{P}(x)\) such that

\[
\Delta_{AB(P)}^{\mathrm{Marg}}
\neq
\mathbb{E}_{X \sim P}\!\left[\Delta_{AB}^{\mathrm{Cond}}(X)\right].
\]

Hence \(\log(\mathrm{RR})\) is \textbf{not directly collapsible in general}.

\vspace{2mm}

\noindent \textbf{Proof (counterexample) of Proposition B2.}

\noindent Let \(X \in \{ 0,1\}\) and \(P(X = 0) = P(X = 1) = 1/2\).

Define conditional means:

\[
\mu_{A}(X = 0) = 0.1,
\quad
\mu_{A}(X = 1) = 0.9,
\qquad
\mu_{B}(X = 0) = 0.2,
\quad
\mu_{B}(X = 1) = 0.9.
\]

Compute the population-average conditional \(\log(\mathrm{RR})\):

\[
\mathbb{E}\!\left[\Delta_{AB}^{\mathrm{Cond}}(X)\right]
=
\frac{1}{2}\log\!\left(\frac{\mu_{B}(0)}{\mu_{A}(0)}\right)
+
\frac{1}{2}\log\!\left(\frac{\mu_{B}(1)}{\mu_{A}(1)}\right)
=
\frac{1}{2}\log\!\left(\frac{0.2}{0.1}\right)
+
\frac{1}{2}\log\!\left(\frac{0.9}{0.9}\right)
=
\frac{1}{2}\log(2).
\]

Compute the marginal \(\log(\mathrm{RR})\):

\[
\Delta_{AB(P)}^{\mathrm{Marg}}
=
\log \frac{\mathbb{E}\!\left[\mu_{B}(X)\right]}
{\mathbb{E}\!\left[\mu_{A}(X)\right]}
=
\log\!\left(
\frac{\tfrac{1}{2}(0.2 + 0.9)}
{\tfrac{1}{2}(0.1 + 0.9)}
\right)
=
\log\!\left(
\frac{0.55}{0.5}
\right)
=
\log(1.1).
\]

Since \(\frac{1}{2}\log(2) \neq \log(1.1)\), we have

\[
\Delta_{AB(P)}^{\mathrm{Marg}}
\neq
\mathbb{E}\!\left[\Delta_{AB}^{\mathrm{Cond}}(X)\right].
\]

Therefore, \(\log( \mathrm{RR}_{AB})\) is not directly collapsible in
general. \(\blacksquare\)

\vspace{2mm}

\noindent \textbf{Proposition B3 (Contrast-induced direct collapsibility of}
\(\mathbf{\log}\left( \mathbf{R}\mathbf{R}_{\mathbf{BC}} \right)\)
\textbf{under (A1) with log link + SEMA).}

\noindent Assume the structure in (A1) with \( g = \log(\cdot) \) and SEMA on the \( g \)-scale:

\[
g\!\left(\mathbb{E}\!\left[Y^{t} \mid X = x\right]\right)
=
m(x)  + \phi(x)+ \delta_{t}\,\mathbf{1}\{t \neq A\},
\]

with \(\delta_{A} \equiv 0\), and the same \(\phi(x)\) for both active
treatments \(B\) and \(C\).

\noindent Then, for every population \(P\),

\[
\Delta_{BC(P)}^{\mathrm{Marg}}
=
\mathbb{E}_{X \sim P}\!\left[\Delta_{BC}^{\mathrm{Cond}}(X)\right]
=
\delta_{B} - \delta_{C},
\]

so the log(RR) for \(B\) vs \(C\) exhibits equivalence
of the marginal and population-average conditional effects. That is,
direct collapsibility is induced (and hence the effect is directly
transportable) under these conditions.

\vspace{2mm}

\noindent \textbf{Proof of Proposition B3.}

\noindent Under (A1),

\[
\eta_{B}(x)
=
m(x)  + \phi(x) + \delta_{B},
\qquad
\eta_{C}(x)
=
m(x)  + \phi(x) + \delta_{C}.
\]

With scale alignment \(h \circ g^{- 1} = I(\cdot)\), (A2) reduces to

\[
\Delta_{BC}^{\mathrm{Cond}}(x)
=
\eta_{B}(x) - \eta_{C}(x)
=
\delta_{B} - \delta_{C},
\]

which is constant in \(x\). Hence

\[
\mathbb{E}_{X \sim P}\!\left[\Delta_{BC}^{\mathrm{Cond}}(X)\right]
=
\delta_{B} - \delta_{C},
\quad \text{for any } P.
\]

Now compute the marginal contrast:

\[
\Delta_{BC(P)}^{\mathrm{Marg}}
=
\log\!\left(
\frac{
\mathbb{E}_{X \sim P}\!\left[\mu_{B}(X)\right]
}{
\mathbb{E}_{X \sim P}\!\left[\mu_{C}(X)\right]
}
\right)
=
\log\!\left(
\frac{
\mathbb{E}_{X \sim P}\!\left[\exp\!\left(\eta_{B}(X)\right)\right]
}{
\mathbb{E}_{X \sim P}\!\left[\exp\!\left(\eta_{C}(X)\right)\right]
}
\right).
\]

Substitute \(\eta_{B},\eta_{C}\):

\[
\begin{aligned}
\mathbb{E}\!\left[\exp\!\left(\eta_{B}(X)\right)\right]
&=
\mathbb{E}\!\left[\exp\!\left(m(X) + \delta_{B} + \phi(X)\right)\right] \\
&=
\exp(\delta_{B})\,
\mathbb{E}\!\left[\exp\!\left(m(X) + \phi(X)\right)\right], \\[10pt]
\mathbb{E}\!\left[\exp\!\left(\eta_{C}(X)\right)\right]
&=
\mathbb{E}\!\left[\exp\!\left(m(X) + \delta_{C} + \phi(X)\right)\right] \\
&=
\exp(\delta_{C})\,
\mathbb{E}\!\left[\exp\!\left(m(X) + \phi(X)\right)\right].
\end{aligned}
\]

Therefore

\[
\begin{aligned}
\Delta_{BC(P)}^{\mathrm{Marg}}
&=
\log\!\left(
\frac{
\exp(\delta_{B})\,
\mathbb{E}\!\left[\exp\!\left(m(X)+\phi(X)\right)\right]
}{
\exp(\delta_{C})\,
\mathbb{E}\!\left[\exp\!\left(m(X)+\phi(X)\right)\right]
}
\right) \\
&=
\log\!\left(\exp(\delta_{B}-\delta_{C})\right)
=
\delta_{B}-\delta_{C}.
\end{aligned}
\]

Combining,

\[
\Delta_{BC(P)}^{\mathrm{Marg}}
=
\delta_{B} - \delta_{C}
=
\mathbb{E}_{X \sim P}\!\left[\Delta_{BC}^{\mathrm{Cond}}(X)\right].
\]

Thus, under (A1) with a log link and SEMA, the B vs C log risk ratio is
directly collapsible. \(\blacksquare\)

\hypertarget{appendix-c.-additional-results-for-illustrative-examples}{%
\subsection{Appendix C. Additional Results for Illustrative
Examples}\label{appendix-c.-additional-results-for-illustrative-examples}}

Building upon the simulated examples presented in Section 7, Appendix C
presents additional results and findings. The main report focused on the
bias plots. Here we also present the effects plot (Figure C1-C4).
Additionally, we present an additional simulation demonstrating that the
log risk ratio is directly transportable under SEMA and scale alignment.
The set-up is similar to the logistic model in Section 7.2, but we fit a
log-binomial model using a log link function and set parameters
\(\beta_{0} = - 3\), \(\beta_{1} = 0.8\), \(\delta_{B} = \log(1.40)\),
\(\delta_{C} = log(1.10)\). Under the SEMA, we set
\(\beta_{2,B} = \beta_{2,C} = - 0.6\), while under the alternative
scenario without SEMA, \(\beta_{2,B} = - 0.6\) and
\(\beta_{2,C} = - 1\).

\vspace{3mm}

\noindent \textbf{Figure C1.} Mean Differences

\noindent \includegraphics[width=1\linewidth]{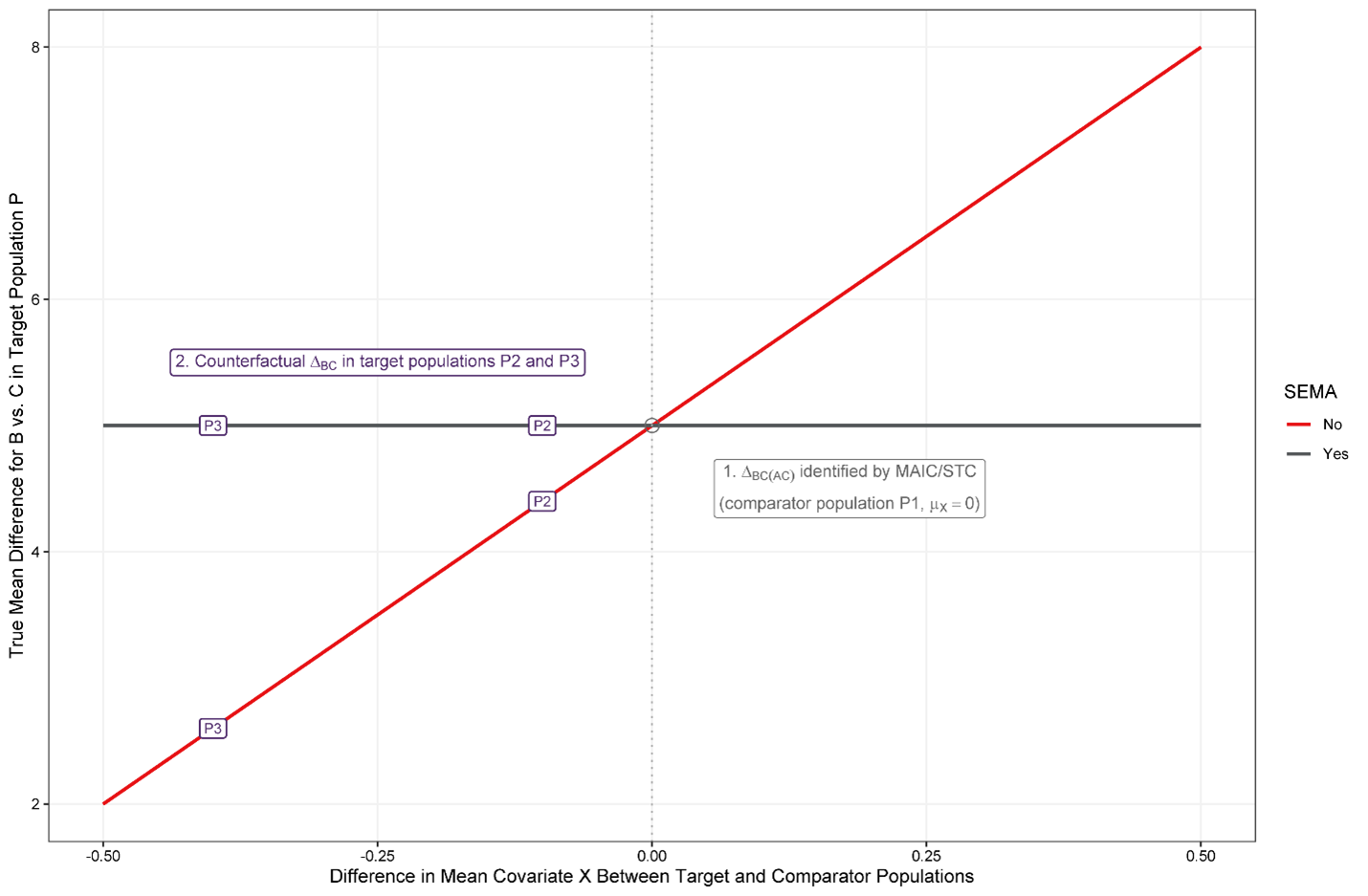}

{\footnotesize
\setlength{\baselineskip}{0.9\baselineskip}
\noindent Flat line indicates direct transportability, whereas a sloped line
indicates Step-2 transport bias.Under SEMA, the mean difference is invariant to changes in the covariate distribution, indicating direct transportability. When SEMA is violated,
the effect varies with the target population, demonstrating failure of direct transportability.\par}

\vspace{3mm}

\noindent \textbf{Figure C2.} Log Odds Ratios

\noindent \includegraphics[width=1\linewidth]{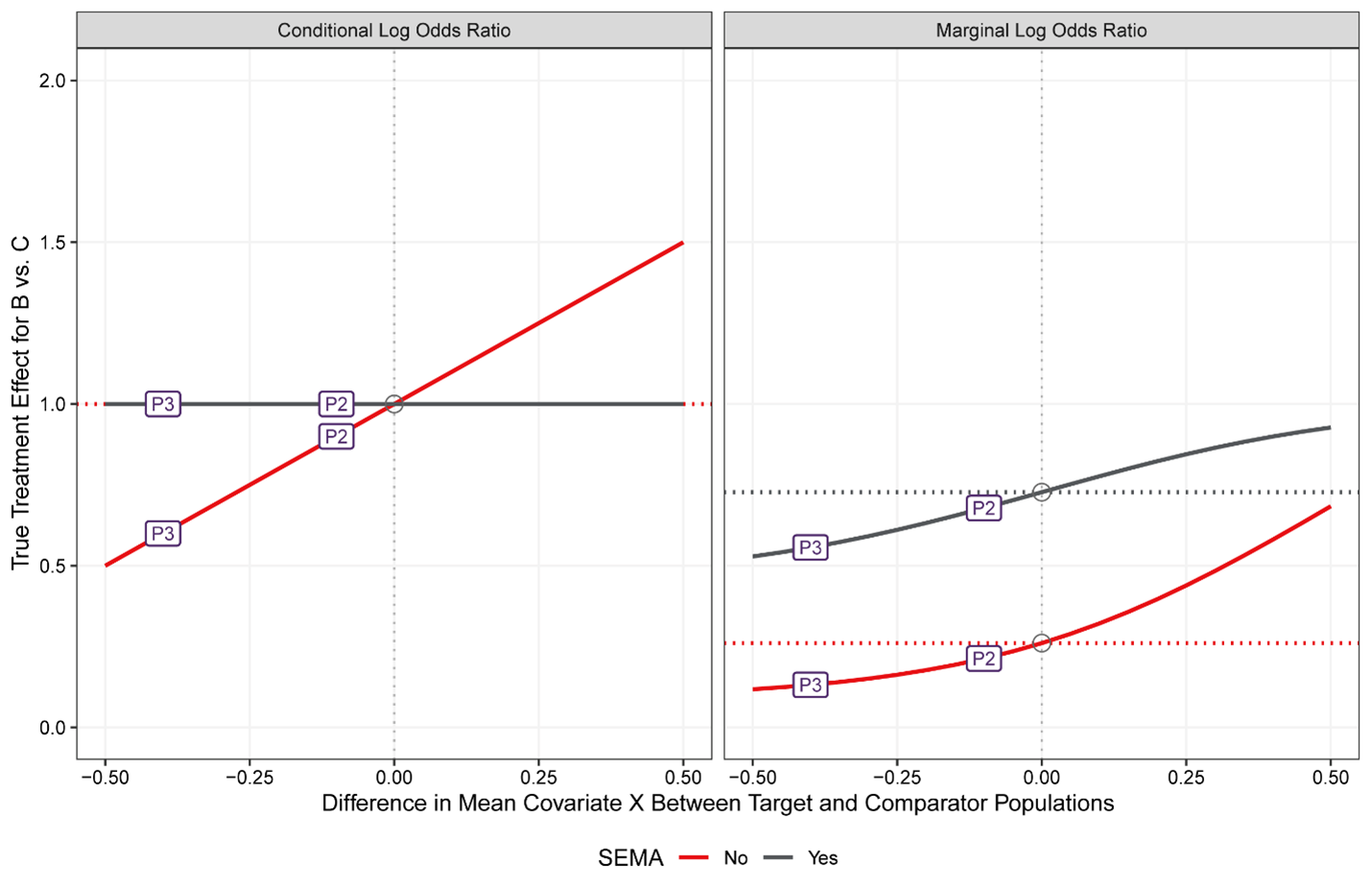}

{\footnotesize
\setlength{\baselineskip}{0.9\baselineskip}
\noindent Conditional (left) and marginal (right) log odds ratios for B vs C
across target populations. Under SEMA, the conditional log odds ratio is
invariant, reflecting direct transportability on the linear predictor
scale. In contrast, the marginal log odds ratio varies with the
covariate distribution even when SEMA holds, illustrating
non-collapsibility and failure of direct transportability.\par}

\vspace{3mm}

\noindent \textbf{Figure C3.} RMST Differences and Ratios

\noindent \includegraphics[width=1\linewidth]{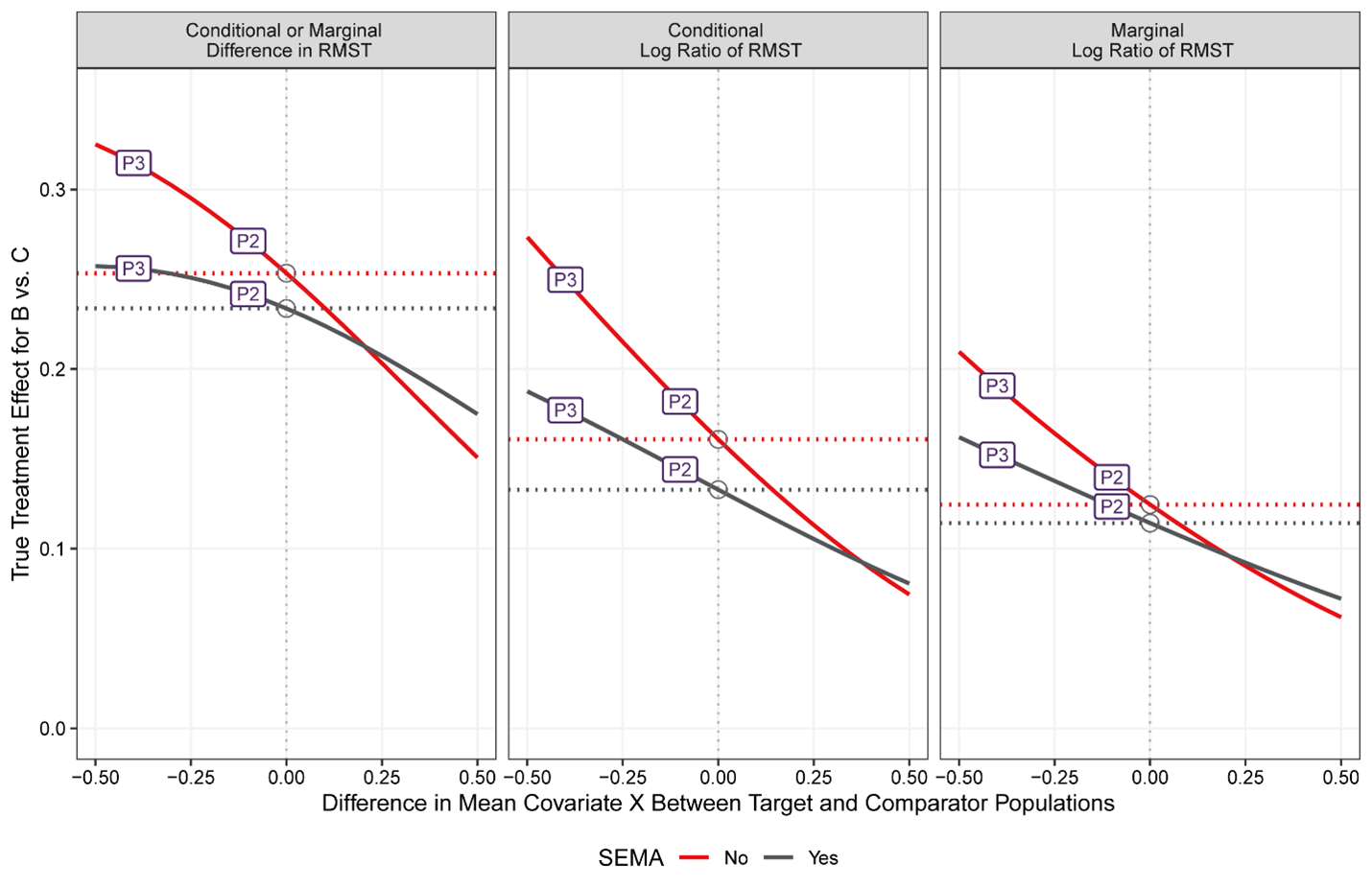}

{\footnotesize
\setlength{\baselineskip}{0.9\baselineskip}
\noindent Conditional and marginal RMST-based contrasts for B vs C across target
populations. Despite RMST differences being directly collapsible within
populations, both RMST differences and ratios vary across populations
under SEMA, reflecting scale misalignment between effect modification
defined on the log-hazard scale and RMST defined on the time scale. This
demonstrates failure of direct transportability due to scale
misalignment rather than non-collapsibility.\par}

\vspace{3mm}

\noindent The direct transportability patterns observed in Figure C4 and C5 for
the log risk ratio reflect the induced direct collapsibility of the
anchored active--active contrast discussed in Section 6.2, where
\(\Delta_{BC}\) becomes invariant to the covariate distribution under
SEMA and scale alignment despite the absence of direct collapsibility
for the individual contrasts against the common comparator.

\vspace{3mm}

\noindent \textbf{Figure C4.} Log Risk Ratios

\noindent \includegraphics[width=1\linewidth]{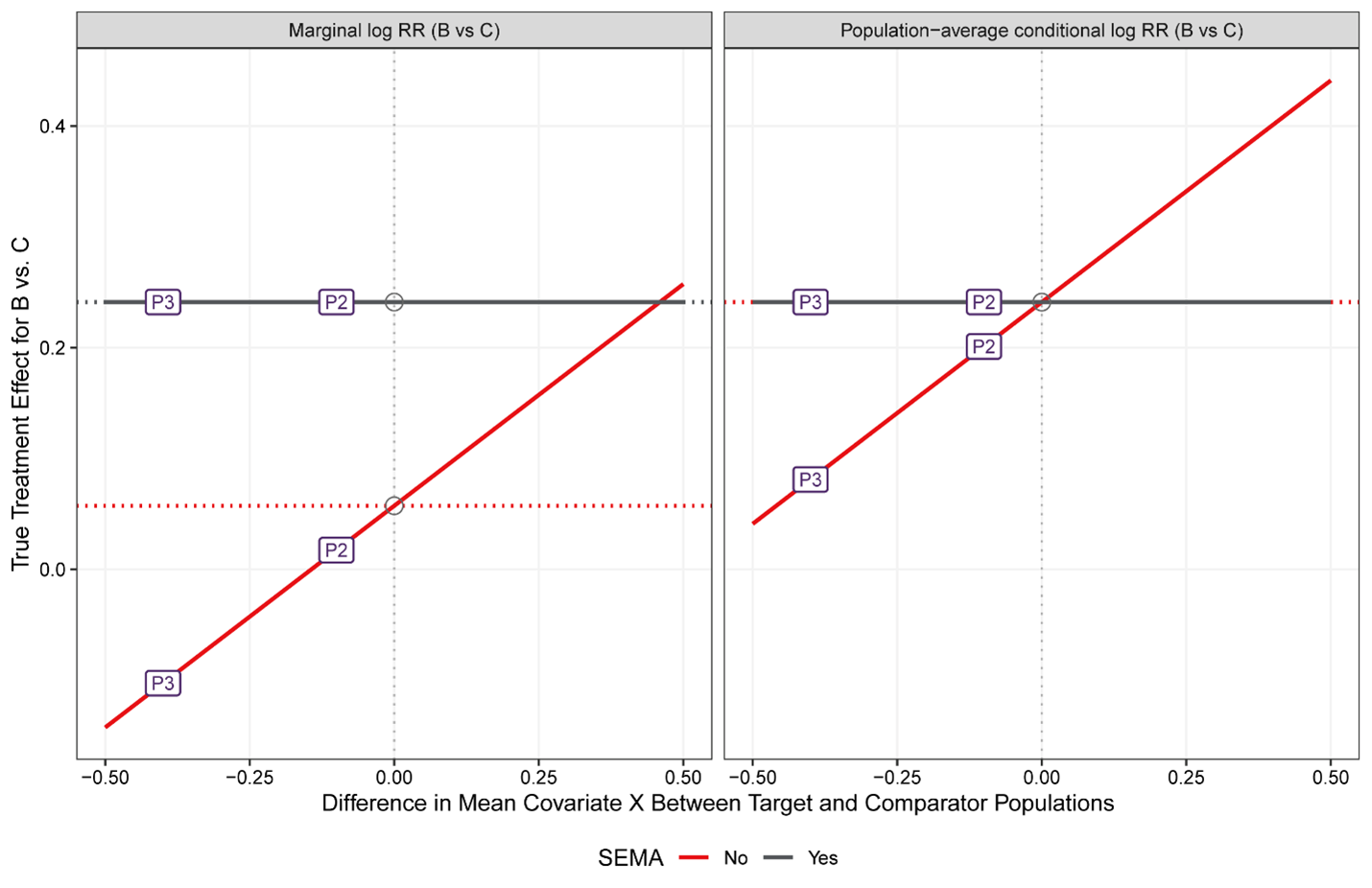}

{\footnotesize
\setlength{\baselineskip}{0.9\baselineskip}
\noindent Marginal and population-average conditional log risk ratios for B vs C
across target populations. When SEMA holds and both effect modification
and the estimand are defined on the log link scale, the log risk ratio
is invariant to changes in the covariate distribution, indicating direct
transportability. When SEMA is violated, the effect varies across
populations. \par}

\vspace{3mm}

\noindent \textbf{Figure C5.} Bias of Directly Transporting Log Risk Ratios

\noindent \includegraphics[width=1\linewidth]{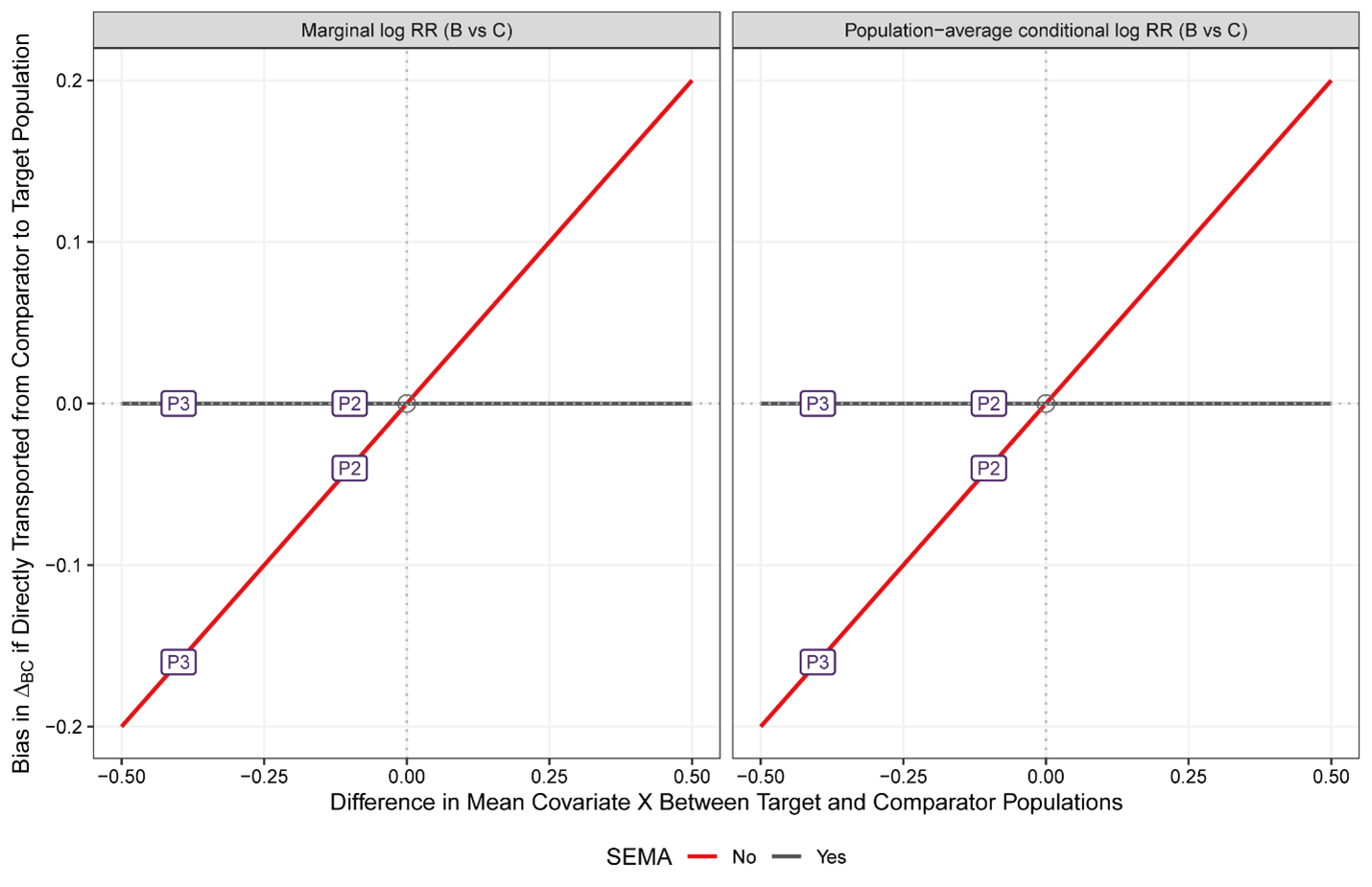}

{\footnotesize
\setlength{\baselineskip}{0.9\baselineskip}
\noindent Under SEMA and scale alignment, no Step-2 transport bias is observed.
When SEMA is violated, bias increases with divergence in the covariate
distribution, confirming that direct transportability relies jointly on
SEMA and scale alignment. \par}

\hypertarget{appendix-d.-population-dependence-of-anchored-marginal-treatment-effects}{%
\subsection{Appendix D. Population Dependence of Anchored Marginal
Treatment
Effects}\label{appendix-d.-population-dependence-of-anchored-marginal-treatment-effects}}

This appendix characterizes which features of the covariate distribution
determine anchored marginal active--active contrasts under common effect
measures. It complements the transportability and collapsibility results
in Appendices A and B by making explicit when dependence reduces to
marginal covariate moments (e.g., means) versus requiring the full joint
distribution, extending insights from Remiro-Azócar (2025) to anchored
PAIC estimands. Note, rate differences and ratios are collapsible when marginal rates are standardized over a common exposure-time distribution across treatment groups. When exposure-time distributions differ between treatments, as with conventional person-time-weighted incidence rates in epidemiology, these measures are generally non-collapsible.

\vspace{2mm}

\noindent \textbf{Table D1.} Population Dependence of Marginal \(\Delta_{BC}\) Estimand

\noindent \includegraphics[
    scale=.73,
    trim=0.1cm 5.5cm 0.1cm 0.1cm,
    clip
]{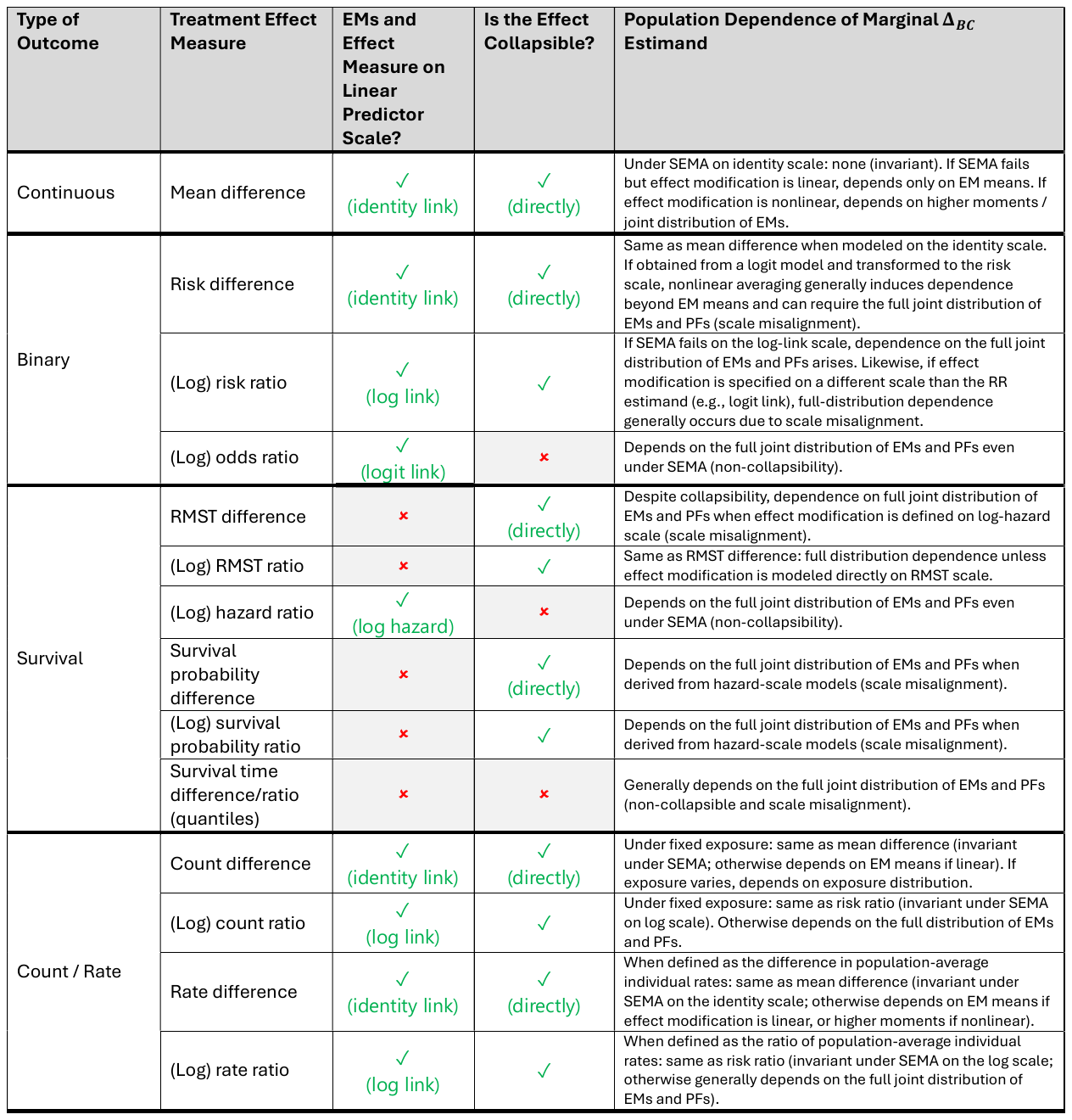}

{\footnotesize
\setlength{\baselineskip}{0.9\baselineskip}
\noindent Reductions to dependence on covariate means assume linear effect
modification on the relevant model (linear predictor) scale, or equivalent conditions allowing \(\mathbb{E}\!\left[\phi(X)\right] = \phi\!\left(\mathbb{E}[X]\right)\). For non-collapsible marginal measures (e.g., OR, HR), marginal \(\Delta_{BC}\) generally depends on the full joint distribution of EMs and PFs even under SEMA; for scale-misaligned estimands (e.g., RMST under log-hazard SEMA), full-distribution dependence arises despite collapsibility. \par}

\hypertarget{d.1-collapsibility-vs.-transportability}{%
\subsubsection{D.1 Collapsibility vs.
Transportability}\label{d.1-collapsibility-vs.-transportability}}

It is important to distinguish between \emph{collapsibility} and
\emph{transportability}, which are related but conceptually distinct
properties. Collapsibility concerns how covariate-specific (conditional)
treatment effects aggregate to marginal effects \emph{within a fixed
population}, and is therefore a property of averaging over covariates
under a single, fixed covariate distribution. Transportability, by
contrast, concerns whether---and under what conditions---a treatment
effect estimand learned in one population can be validly applied to
another population with a different covariate distribution, which
entails re-expressing or re-averaging the estimand under a new covariate
distribution. In the special case of \emph{direct transportability},
this cross-population re-averaging leaves the estimand invariant; more
generally, \emph{conditional transportability} requires explicit
re-standardization to the target population. From the scale-aligned
estimand perspective adopted here, this distinction can be understood in
terms of where averaging over covariates occurs relative to the
transformation \(h( \cdot )\) defining the estimand and whether effect
modification is additive on the scale of the linear predictor defined by
\(g( \cdot )\). Specifically, collapsibility concerns averaging after
application of \(h( \cdot )\) within a population, whereas direct
transportability depends on whether re-averaging under a different
covariate distribution preserves the estimand across populations. As a
result, even effect measures that are collapsible within populations may
fail to be directly transportable across populations when the scale on
which effect modification is defined does not align with the estimand
scale.

In this sense, collapsibility is a \emph{within-population} property of
an effect measure, whereas transportability is a \emph{cross-population}
property of an estimand.

\hypertarget{appendix-e.-paic-methods-for-achieving-conditional-transportability}{%
\subsection{Appendix E. PAIC Methods for Achieving Conditional
Transportability}\label{appendix-e.-paic-methods-for-achieving-conditional-transportability}}

PAIC methods are designed to adjust for cross-study differences in
baseline characteristics when PLD are available for only a subset of
studies. Their goal is to estimate relative treatment effects within a
common target population by reweighting, modeling, or integrating
information across trials. There are three methods for PAICs: 1)
matching-adjusted indirect comparisons (MAICs), 2) simulated treatment
comparisons (STCs), and 3) multilevel network meta regression (ML-NMR).
Each method differs in how it models treatment effect heterogeneity,
defines the target population, and implements direct or conditional
transportability.

Briefly, MAIC estimates balancing weights using the method of moments
and makes use of Equation 12 to conditionally transport marginal effects
from the population of the index study to the comparator study. MAIC
inherently targets marginal effects because it reweights observed
outcomes rather than modeling conditional relationships. On the other
hand, STCs and ML-NMR can target both marginal and conditional effects.
Both approaches are formulated using outcome regression models, but
their transportability properties differ, as explored in the subsequent
sections. When targeting marginal effects, STC conditionally transports
estimates from the population of the index study to the comparator study
using g-computation, as defined in Equation 11. ML-NMR extends Equation
11 to accommodate multiple arms and studies (Equation 26), further
discussed in Sections 9.1-9.2. This extension enables results to be
conditionally transported to any target population under additional
modeling assumptions. To date, ML-NMR has only been used in the anchored
setting. We introduce an extension of ML-NMR for unanchored comparisons
in Section 9.3.

While these methods differ in complexity and scope, in practice,
pairwise applications of MAICs and STCs remain the most widely used in
HTA submissions. This appendix summarizes PAIC methods specifically to
clarify how each implicitly defines a source population, target
population, and estimand, which is central to the two-step
transportability arguments developed in Sections 4--7.

\vspace{2mm}

\noindent \textbf{Table E1.} Statistical Methods for Achieving Conditional Transportability

\setlength{\tabcolsep}{4pt}
{\footnotesize
\begin{longtable}{@{}L{0.23\linewidth} L{0.23\linewidth} L{0.23\linewidth} L{0.23\linewidth}@{}}
\toprule
\textbf{Criterion} & \textbf{MAIC} & \textbf{STC} & \textbf{ML-NMR} \\
\midrule
\endfirsthead

\toprule
\textbf{Criterion} & \textbf{MAIC} & \textbf{STC} & \textbf{ML-NMR} \\
\midrule
\endhead

\midrule
\multicolumn{4}{r}{\footnotesize Continued on next page} \\
\midrule
\endfoot

\bottomrule
\endlastfoot

Source population & 1 index trial & 1 index trial & All index and comparator studies \\
Target population & Comparator trial & Comparator trial & Any (user-defined, including real-world or HTA-relevant) \\
Conditional transportability mechanism &
Reweighting (method of moments) &
Outcome regression + g-computation &
Hierarchical outcome regression (multilevel model) \\
Target estimand &
Marginal &
Marginal (conditional is theoretically possible, but limited by comparator data availability) &
Marginal or conditional \\
SEMA &
Not required unless direct transport is invoked in Step 2 &
Not required unless direct transport is invoked in Step 2 &
Required for transportability to any population \\
Anchor vs.\ unanchored &
Anchored and unanchored &
Anchored and unanchored &
Anchored (extensions allow unanchored; see Section 9.3) \\
Handles multiple comparators / studies & No & No & Yes \\
\end{longtable}
}

\end{document}